\documentclass[prd,aps,
tightenlines,
superscriptaddress,showpacs,nofootinbib,eqsecnum,amsfonts,amsmath
]{revtex4}
\usepackage{bm,graphics,graphicx,epsfig,color
,ulem}
\allowdisplaybreaks
\def\be{\begin{equation}}
\def\ee{\end{equation}}

\def\gsim{\mathrel{
\rlap{\raise 0.511ex \hbox{$>$}}{\lower 0.511ex
\hbox{$\sim$}}}}
\def\lsim{\mathrel{
\rlap{\raise 0.511ex \hbox{$<$}}{\lower 0.511ex
\hbox{$\sim$}}}}
\begin{document}
\title{Head-on infall of two compact objects: Third post-Newtonian Energy Flux} 
\author{Chandra Kant Mishra}\email{chandra@rri.res.in}
\affiliation{Raman Research Institute, Bangalore
560 080, India}
\affiliation{Indian Institute of Science,Bangalore 560 012, India}
\author{Bala R. Iyer} \email{bri@rri.res.in} \affiliation{Raman Research
Institute, Bangalore 560 080, India}
\date{\today}
\pacs{PACS numbers: 04.25.Nx, 04.30.Db, 97.60.Jd, 97.60.Lf}
\begin{abstract}
Head-on infall of two compact objects with arbitrary mass ratio is investigated using the multipolar post-Minkowskian approximation method. At the third post-Newtonian order the energy flux, in addition to the instantaneous contributions, also includes hereditary contributions consisting of the gravitational-wave tails, tails-of-tails and the tail-squared terms. The results are given both for infall from infinity and also for infall from a finite distance. These analytical expressions should be useful for the comparison with the high accuracy numerical relativity results within the limit in which post-Newtonian approximations are valid.
\end{abstract}
\maketitle
\section{Introduction}
\label{intro}  
The spiraling coalescence of two compact objects (black holes or neutron stars) moving about one another in an orbit, forms a prominent class of sources of gravitational radiation~\cite{1994reco.conf...67T}. Such sources of gravitational waves (GW), especially in their late stages of evolution are prime targets for gravitational wave detectors such as LIGO~\cite{Science92} and Virgo~\cite{bss:virgo}. The evolution of the binary systems composed of two compact objects involves three stages of evolution; the early inspiral, late inspiral and merger and the final ringdown. Detection of gravitational radiation from such systems by the gravitational wave detectors depends strongly on the theoretical inputs, which will involve  computation of the  waveform of the signal for all the three phases to very high post-Newtonian (PN) order to detect and infer the characteristics of the sources of GWs, using matched filtering techniques~\cite{Th300}. Even though  head-on collision  of two black holes
has only a small astrophysical possibility, it provides the simplest possible situation to study the two-body problem of general relativity and has been  studied since it provides an excellent
theoretical  platform for comparing the validity of various analytical and numerical approaches towards solving Einstein's equations in dynamical situations.

One of the earliest attempts to solve the problem of head-on collision using a complete general relativistic approach was due to Davis et al~\cite{DRPP71}. They discussed the emission of gravitational radiation due to the radial infall of a test particle in Schwarzschild spacetime from infinity, using Zerilli's equation for black-hole perturbations~\cite{Zerilli70}. Because of the axial symmetry of the system, the problem simplifies considerably and yet retains the features of astrophysical interest such as emission of gravitational radiation at infinity. In addition to this, head-on collision can be considered as an approximation to the last stage of the inspiralling coalescence-when two objects merge together to form a single object. The first attempt to solve the head-on collision of two equal mass black holes numerically was due to Smarr and Eppley~\cite{SmarrThesis, EppleyThesis, Smarr79}. This program has undergone substantial improvement in accuracy and reliability with advances in the understanding of numerical issues in the treatment of Einstein's equations and availability of better computing~\cite{Anninos:1993zj}. The head-on collision of two black holes with arbitrary mass ratio has been investigated numerically in~\cite{Anninos:1998wt,Choptuik:2009ww}
and semianalytically~\cite{Berti:2010ce}.
In a recent work~\cite{Sperhake:2008ga} head-on collision of two equal mass, nonrotating black holes with ultrarelativistic speeds have been studied using numerical methods. The main result of this
analysis is that in such a process (where the initial energy of the system is dominated by kinetic energy of black holes) the total amount of energy converted to gravitational waves is about 14\% of the initial mass-energy for the system
and corresponds to large luminosities  of the order of $10^{-2}\,c^5/G$. Another study related to the collision of two equal mass, nonrotating black holes moving at ultrarelativistic speeds and with generic impact parameter~\cite{Sperhake:2009jz} suggests that such collisions can produce black holes rotating close to the Kerr limit and the energy radiated in such a process would be roughly 35\% of the center-of-mass (CM) energy.      

Another approach which may be used to study the head-on collision of two compact objects is the PN approximation approach. Though PN methods are valid for arbitrary mass ratios, they eventually break down under situations like strong gravitational fields and high speeds. Simone, Poisson and Will (SPW)~\cite{SPW95} investigated the problem of head-on infall and compared the PN approach with black-hole perturbation (BHP) theory. They provided  2PN accurate expression for the far-zone GW energy flux and showed, in particular, that the
energy radiated during the infall is wellestimated by the quadrupole approximation combined with the  \textit{exact} test-body equations of motion (EOM)
in Schwarzschild background. Also in a recent study~\cite{Nichols:2010qi}, a hybrid method using both PN approximations and BHP theories has been used to study the head-on collision of two black holes and found that PN and BHP theories can explain the main features of gravitational radiation for head-on mergers.

In this paper we investigate the problem of head-on infall using the multipolar post-Minkowskian (MPM) approach~\cite{Bliving,BFeom,BIJ02,BI04mult,BDE04,BDEI05} and provide the complete 3PN accurate expression for the GW energy flux emitted during the radial infall of two compact objects towards each other. In addition to the simpler instantaneous part of the energy flux we also compute the more complex hereditary contributions up to 3PN order which involves the contributions due to tails, tails-of-tails and tail-squared terms. We discuss the head-on problem  both for  infall starting from rest at an initial finite separation (denoted case I) and  similarly for infall starting from rest at infinite separation (denoted by case II). 
Instantaneous contributions at 2.5PN order and at 3PN order, computation of tails at 2.5PN, tail-of-tail and tail-squared terms at 3PN order are the  new results of this paper.
Our computations suggest that the total energy radiated in the process of head-on infall of two compact objects with equal masses is roughly about 0.0074\% of the Arnowitt, Deser, and Misner (ADM) mass of the binary and the peak luminosities are typically less than of the order  $5\times 10^{-6}c^5/G$. Comparing our PN estimates with the  numerical relativity results~\cite{Anninos:1993zj} we can see that 
the PN estimates are smaller than the numerical results typically by a factor of $27$ consistent with the expectation that
 a larger fraction of energy radiated indeed
comes from the merger phase of the infall rather than from the early inspiral.

This paper is organized in the following way. In Sec.~\ref{structure-GWEF} we begin by providing the structure of the far-zone GW energy flux at 3PN order, relations connecting radiative multipole moments to source multipole moments and the decomposition of the expression for energy flux into instantaneous and hereditary contributions. 
Section~\ref{EqnOM} lists the 3PN EOM as well as the 3PN accurate expression for the center-of-mass energy in standard harmonic coordinates for the head-on case. 
In Sec.~\ref{sourcemoments} we give the expressions for the desired multipole moments at the PN order required for the computation of 3PN energy flux for head-on infall case. 
In Sec.~\ref{3PNEFINST} we first exhibit the instantaneous part of energy flux up to 3PN order in standard harmonic coordinates
followed by  the corresponding expressions in two alternative coordinates 
for possible comparison with numerical relativity results: modified harmonic (MH) and ADM.
 Section~\ref{hereditary} describes the computation of the hereditary part of the energy flux. Finally, in Sec.~\ref{completeflux}, we bring together the complete 3PN accurate expression for energy flux in ADM coordinates and the energy radiated during infall to some fixed radial coordinate.
Section~\ref{discussion} contains a graphical display of the  salient features and our conclusions. 
These results should be useful to compare and match to simulations using numerical methods in regimes where both treatments
are expected to be the valid.         
The paper ends with a short appendix relating the expression for conserved energy in standard harmonic (SH) coordinates to that in ADM coordinates. 
\section{The Far-zone GW Energy Flux}
\label{structure-GWEF}
We start the discussion by writing the 3PN expression for far-zone GW energy flux in terms of the symmetric trace-free radiative multipole moments~\cite{Th80, ABIQ07}. The PN structure for GW energy flux reads as,
\begin{align}
\mathcal{F}(U) &={G\over c^5}\Biggl\{ {1\over 5}\, U^{(1)}_{ij}
U^{(1)}_{ij}\nonumber\\ &\quad+{1\over c^2} \left[ {1\over 189}
\,U^{(1)}_{ijk} U^{(1)}_{ijk} +{16\over 45}\, V^{(1)}_{ij}
V^{(1)}_{ij}\right]+{1\over c^4} \left[ {1\over 9072}\, U^{(1)}_{ijkm}
U^{(1)}_{ijkm}+{1\over 84}\, V^{(1)}_{ijk}
V^{(1)}_{ijk}\right]\nonumber\\ &\quad+{1\over c^6}\left[ {1\over
594000}\, U^{(1)}_{ijkmn} U^{(1)}_{ijkmn}+{4\over 14175}\, V^{(1)}_{ijkm}
V^{(1)}_{ijkm}\right]+\mathcal{O}(8)\Biggr\}\,.
\label{EFUV}
\end{align}
In the above expression $U_{L}$ and $V_{L}$(where $L=i_1i_2\cdots i_l$
represents a multi-index composed of $l$ spatial indices) are the mass-type and current-type radiative multipole moments respectively and
$U_{L}^{(n)}$ and $V_{L}^{(n)}$ denote their $n^\mathrm{th}$ time
derivatives. The moments appearing in the formula are functions of retarded time $U\equiv
T-{\it R}/c$ in radiative coordinates.    

Equation~\eqref{EFUV} is the general formula for the computation of 3PN accurate energy flux for any general isolated source. In a recent paper~\cite{ABIQ07} the complete third post-Newtonian energy flux has been computed for inspiralling compact binaries moving in quasi-elliptical orbits. In the present work we specialize to the case of head-on infall and compute the 3PN accurate far-zone GW energy flux emitted due to head-on infall  of two compact objects with arbitrary mass ratio using  the MPM approximation method. 
The radiative current-type moments ($V_L$) are related to the source current
moments $J_L$ whose expansion at each PN order contains the orbital
angular momentum ${\cal {J}}$ which vanishes in the head-on case. 
Thus the current-type moments $V_L$ will not contribute to GW energy flux 
and for the head-on case, Eq.~\eqref{EFUV} reduces to the following form,      
\begin{align}
\mathcal{F}(U) &={G\over c^5}\Biggl\{ {1\over 5}\, U^{(1)}_{ij}
U^{(1)}_{ij}
+{1\over c^2} \left[ {1\over 189}\,
U^{(1)}_{ijk} U^{(1)}_{ijk}\right]+{1\over c^4} \left[ {1\over 9072}\, U^{(1)}_{ijkm}
U^{(1)}_{ijkm}\right]
+{1\over c^6}\left[ {1\over
594000}\, U^{(1)}_{ijkmn} U^{(1)}_{ijkmn}\right]+\mathcal{O}(8)\Biggr\}\,.
\label{EFUVHOC}
\end{align}

In the MPM formalism, the radiative moments $U_L$ and $V_L$ are related to canonical moments $M_L$ and $S_L$ respectively
and these canonical moments are in turn expressed in terms of source moments $\{I_L,J_L,W_L,X_L,Y_L,Z_L\}$. Since in the present work we only deal with head-on situation we would exclude terms involving current-type multipole moments from all our expressions for the reason stated above. It should be evident from the Eq.~\eqref{EFUVHOC} that for the computation of 3PN accurate energy flux $U_{ij}$ is needed at 3PN order, $U_{ijk}$ is needed at 2PN order, $U_{ijkl}$ with 1PN accuracy and $U_{ijklm}$  to leading Newtonian accuracy. General expressions connecting $U_{L}$ to source moments have been listed in~\cite{ABIQ07} and we shall simply recall those expressions. For the 3PN accurate mass quadrupole we have      
\begin{align}
U_{ij}(U) &= I^{(2)}_{ij} (U) + {2\,G\,M\over c^3} \int_{0}^{+\infty} d
\tau \left[\log \left({c\,\tau \over 2\,r_0}\right)+{11\over 12} \right]
I^{(4)}_{ij} (U-\tau) \nonumber \\
&+\frac{G}{c^5}\left\{-\frac{2}{7}\int_{0}^{+\infty} d\tau
I^{(3)}_{a\langle i}(U-\tau)I^{(3)}_{j\rangle a}(U-\tau) \right.\nonumber \\
&\qquad~ + {1 \over7}\,I^{(5)}_{a\langle i}I_{j\rangle a} - {5 \over7}\,
I^{(4)}_{a\langle i}I^{(1)}_{j\rangle a} -{2 \over7}\, I^{(3)}_{a\langle i}I^{(2)}_{j\rangle a}
\nonumber\\ &\qquad~\left.
+4\left[W^{(2)}I_{ij}-W^{(1)}I_{ij}^{(1)}\right]^{(2)} \right\}\nonumber \\
&+2\left(\frac{G\, M}{c^3}\right)^2\int_{0}^{+\infty}d\tau I_{ij}^{(5)}
\left(U-\tau\right)\,\left[\log^2\left({c\,\tau \over
2\,r_0}\right)+{57\over70}\, \log\left({c\,\tau \over
2\,r_0}\right)+{124627\over44100}\,\right] +\mathcal{O}(7),\label{Uij}
\end{align}
where the bracket $<>$ surrounding indices denotes the symmetric trace-free projection. The $I_{L}$'s are the mass-type source moments (and $I_{L}^{(n)}$ denote their $n^\mathrm{th}$ time derivatives), and $W$ is the monopole corresponding to the gauge moment $W_L$ which for our purpose needs to be known  Newtonian accuracy. The quantity $M$ appearing in the above expression is the ADM mass of the source. It should be evident from Eq.~\eqref{Uij} that radiative moments have two distinct contributions. The first referred to as the \textit{instantaneous} contribution requires the knowledge of source multipole moments only at a given retarded time, $U=T-{\it R}/c$; where {\it R} is the distance of the source in radiative coordinates. The second one, 
referred to as the \textit{hereditary} contribution, which is given by integrals over retarded time from 0 to $\infty$, depends on the dynamics of the system in its entire past history and requires the knowledge of source moments at all times before $U$.
A closer look at the hereditary terms reveals two types of contributions, some with and some without the $\log$ factors. The integrals (with $\log$ factors) appearing at 1.5PN and 3PN order are called tail and tail-of-tail integrals respectively. The integral (without $\log$ factor) appearing at 2.5PN order is called the nonlinear memory integral. It is a time antiderivative and hence leads to an instantaneous term in the energy flux.

The mass-type octupole moment $U_{ijk}$ which is needed at 2PN is related to the associated source moment as
\begin{align}
U_{ijk} (U) &= I^{(3)}_{ijk} (U) + {2\,G\,M\over c^3} \int_{0}^{+\infty}
d\tau\left[ \log \left({c\,\tau \over 2\,r_0}\right)+{97\over60} \right]
I^{(5)}_{ijk} (U-\tau)+\mathrm{\mathcal{O}(5)} \,.\label{Uijk}
\end{align}
For other radiative moments, $U_{ijkl}$ and $U_{ijklm}$, only the leading order accuracy in the relation between radiative and source moments is needed, so that
\begin{equation}
U_{L}(U)=I^{(l)}_{L}(U)+\mathrm{\mathcal{O}(3)}\,.
\label{UL}
\end{equation}
The constant $r_0$ which provides a scale for the logarithmic term in the above expressions is an arbitrary constant. It enters the relation connecting retarded time $U=T-{\it R}/c$ in radiative coordinates to retarded time $u=t-r/c$ in harmonic coordinates (where $r$ is the distance of the source in harmonic coordinates). The relation between retarded time in radiative coordinates, and the one in harmonic coordinates reads as
\begin{equation}
U=t-{r\over c}-{2\,G\,M \over c^3}\, \log \left({r \over r_0}\right)+\mathrm{\mathcal{O}(5)}\,.
\label{U-u}
\end{equation}
Later in this paper we shall show that the presence of this constant $r_0$ will not influence any physical result like far-zone GW energy flux.

We can now use the expressions for the radiative moments given by Eqs.~\eqref{Uij}-\eqref{UL} in Eq.~\eqref{EFUVHOC} to obtain the 3PN energy flux formula in terms of source moments. As discussed above, the presence of two distinct contributions (instantaneous and hereditary) leads to a natural decomposition of the 3PN energy flux into two pieces and the complete flux can be written as a sum of the two 
distinct types of contributions as
\begin{equation}
\mathcal{F}=\mathcal{F}_{\rm
inst}+\mathcal{F}_\mathrm{hered}\,,
\label{tot}
\end{equation}     
where the instantaneous contribution\footnote{There is a typographical error in Eq.~(2.7) of~\cite{ABIQ07} 
which has been corrected while writing Eq.~(\ref{inst}) of the present work.  At 2.5PN order the coefficient of $I^{(3)}_{ij}I^{(3)}_{ij}$  should be $-{4\over 7}$ 
and not  $-{2 \over 7}$. However, the results in~\cite{ABIQ07} are computed using the correct coefficient.} to the energy flux is given by
\begin{align}
\mathcal{F}_\mathrm{inst}(U)&={G\over c^5}\biggl\{ {1\over 5}\,
I^{(3)}_{ij} I^{(3)}_{ij}
+{1\over c^2} \left[ {1\over
189}\, I^{(4)}_{ijk} I^{(4)}_{ijk}
\right]+{1\over c^4} \left[ {1\over 9072}\, I^{(5)}_{ijkm}
I^{(5)}_{ijkm}
\right]\nonumber\\
&+\frac{G}{c^5}\left[\frac{8}{5}\,I_{ij}^{(3)}\left(I_{ij}W^{(5)}+2
\,I_{ij}^{(1)}W^{(4)}-2\,I_{ij}^{(3)}W^{(2)}-I_{ij}^{(4)}W^{(1)}\right)\right.\nonumber\\
&\qquad\left.+\frac{2}{5} \,I_{i j}^{(3)} \left(-\frac{4}{7}\,
I_{ai}^{(5)} I_{aj}^{(1)}-I_{ai}^{(4)} I_{aj}^{(2)}-\frac{4}{7}
\,I_{ai}^{(3)} I_{aj}^{(3)}+\frac{1}{7}\, I_{ai}^{(6)} I_{aj}
\right)
\right]\nonumber\\ &+{1\over c^6}\left[ {1\over 594000}\,
I^{(6)}_{ijkmn} I^{(6)}_{ijkmn}
\right]+\mathcal{O}(7)\biggr\}\,.
\label{inst}
\end{align}
The hereditary contribution comprises of three parts,
\begin{equation}
\mathcal{F}_\mathrm{hered}=\mathcal{F}_{\rm tail}+\mathcal{F}_{\rm tail(tail)}+\mathcal{F}_{\rm (tail)^2}\,,
\label{hered}
\end{equation}
where the quadratic-order (proportional to $G^2$) tails are given by 
\begin{align}\label{FtailR}
\mathcal{F}_\mathrm{tail}(U) &= \frac{4\,G^2\,M}{5\,c^8}\,I_{ij}^{(3)}(U)
\int_0^{+\infty}d\tau\,I^{(5)}_{ij}(U-\tau)\biggl[\log\left(\frac{c\,\tau}{2\,r_0}\right)
+\frac{11}{12}\biggr]\nonumber\\ &+ \frac{4\,G^2\,M}{189\,
c^{10}}\,I_{ijk}^{(4)}(U)\int_0^{+\infty}d\tau\,I^{(6)}_{
ijk}(U-\tau)\biggl[\log\left(\frac{c\,\tau}{2\,r_0}\right)+\frac{97}{60}\biggr]\,,
\end{align}
and the cubic-order tails (proportional to $G^3$) by
\begin{subequations}
\label{FtailtailR}\begin{align}
\mathcal{F}_\mathrm{tail(tail)}(U) &=
\frac{4\,G^3\,M^2}{5\,c^{11}}\,I_{ij}^{(3)}(U)\int_0^{+\infty}d\tau\,I^{(6)}_{
ij}(U-\tau)\biggl[\log^2\left(\frac{
c\,\tau}{2\,r_0}\right)+\frac{57}{70}\,\log\left(\frac{
c\,\tau}{2\,r_0}\right)+\frac{124627}{44100}\biggr],\\
\mathcal{F}_\mathrm{(tail)^2}(U) &=
\frac{4\,G^3\,M^2}{5\,c^{11}}\left(\int_0^{+\infty}d\tau\,I^{(5)}_{
ij}(U-\tau)\biggl[\log\left(\frac{
c\,\tau}{2\,r_0}\right)+\frac{11}{12}\biggr]\right)^2\,.
\end{align}\end{subequations}    
Here one should note that the general formulae for energy flux include some contributions from current-type moments as well (see~\cite{ABIQ07, ABIQ07tail}) but these vanish for the head-on case. Further, it should be noted that Eqs.~\eqref{Uij}-\eqref{Uijk} and thus Eqs.~\eqref{FtailR}-\eqref{FtailtailR} show an intermediate dependence on the arbitrary length scale $r_0$ which should eventually cancel from all physical quantities. Such a cancellation of the scale $r_0$ from all physical quantities occurs naturally in the  MPM formalism and has been explicitly
shown for  sources such as binary systems moving in circular~\cite{BIJ02} and elliptical orbits~\cite{ABIQ07tail, B98tail}. 
This is facilitated because an explicit computation of the hereditary integrals
is possible since the integral over the complete past in the adiabatic approximation reduces to an integral over  the current (noninspiralling) orbit which can then be  computed  making explicit use of the periodicity features in the motion.
For a head-on situation, on the other hand, the absence of periodicity prevents the straightforward extension of the above method. 
A more first-principle treatment is called for based on the observation that  
since most of the $r_0$- dependence comes from our definition [Eq.~\eqref{U-u}] of a radiative coordinate system, it can be tracked and isolated by inserting $U$ as given by Eq.~\eqref{U-u} back in Eq.~\eqref{Uij}-\eqref{Uijk}. Upon doing so we get expressions for radiative moments in harmonic coordinates ($t, r$), which read
\begin{align}
U_{ij}(u) &= I^{(2)}_{ij} (u) + {2\,G\,M\over c^3} \int_{0}^{+\infty} d
\tau \,\left[ \log \left({c\,\tau \over 2\,r}\right)+{11 \over 12} \right]
I^{(4)}_{ij} (u-\tau) \nonumber \\
&+\frac{G}{c^5}\left\{-\frac{2}{7}\int_{0}^{+\infty} d\tau\,
I^{(3)}_{a\langle i}(u-\tau)I^{(3)}_{j\rangle a}(u-\tau) \right.
\nonumber \\&\qquad~ 
+ {1 \over7}\,I^{(5)}_{a\langle i}I_{j\rangle a} - {5 \over7}\,
I^{(4)}_{a\langle i}I^{(1)}_{j\rangle a} -{2 \over7}\, I^{(3)}_{a\langle i}I^{(2)}_{j\rangle a}\nonumber\\ &\qquad~\left.
+4\left[W^{(2)}I_{ij}-W^{(1)}I_{ij}^{(1)}\right]^{(2)} \right\}\nonumber \\
&+2\left(\frac{G\, M}{c^3}\right)^2\int_{0}^{+\infty}d\tau\, I_{ij}^{(5)}
\left(u-\tau\right)\left[\log^2\left({c\,\tau \over
2\,r}\right)+{57\over70}\, \log\left({c\,\tau \over
2\,r}\right)+{124627\over44100}\right]\nonumber\\ &-{214 \over 105}\, \log \left({r \over r_0}\right)\,\left({G\, M \over c^3}\right)^2\, I^{(4)}_{ij}(u)+\mathcal{O}(7),\label{Uijhar}
\end{align}  
\begin{align}
U_{ijk}(u) &= I^{(3)}_{ijk} (u) + {2\,G\,M\over c^3} \int_{0}^{+\infty}
d\tau\left[ \log \left({c\,\tau \over 2\,r}\right)+{97\over60} \right]
I^{(5)}_{ijk} (u-\tau)+\mathrm{\mathcal{O}(5)} \,.\label{Uijkhar}
\end{align}
In the 1.5PN term, the $r_0$ dependence is more trivial and disappears with
the change from radiative to harmonic coordinates. At 3PN order, however,
there still remains a nontrivial $r_0$ dependent term.
However, the quadrupole mass moment $I_{ij}$ also depends on the constant $r_0$ at 3PN order [see Eq.~\eqref{MQHOC}]  and when one takes those dependences into account we will check that the 3PN radiative moment $U_{ij}$ is indeed independent of $r_0$. 
Using the expressions for radiative moments given by Eqs.~\eqref{Uijhar} and \eqref{Uijkhar} we can rewrite explicitly the different hereditary contributions given by Eqs.~\eqref{hered}, \eqref{FtailR}, and \eqref{FtailtailR}. One finds the quadratic-order(proportional to $G^2$) tails are given by 
\begin{align}\label{Ftailhar}
\mathcal{F}_\mathrm{tail}(u) &= \frac{4\, G^2\,M}{5\,c^8}\,I_{ij}^{(3)}(u)\,
\int_0^{+\infty}d\tau\,I^{(5)}_{ij}(u-\tau)\biggl[\log\left(\frac{c\,\tau}{2\,r}\right)
+\frac{11}{12}\biggr]\nonumber\\ &+ \frac{4\,G^2\,M}{189
\,c^{10}}\,I_{ijk}^{(4)}(u)\,\int_0^{+\infty}d\tau\,I^{(6)}_{
ijk}(u-\tau)\biggl[\log\left(\frac{c\,\tau}{2\,r}\right)+\frac{97}{60}\biggr]\,,
\end{align}
and the cubic-order tails (proportional to $G^3$) by
\begin{subequations}
\label{Ftailtailhar}
\begin{align}
\mathcal{F}_\mathrm{tail(tail)}(u) &=
\frac{4\,G^3\,M^2}{5\,c^{11}}\,I_{ij}^{(3)}(u)\,\int_0^{+\infty}d\tau\,I^{(6)}_{
ij}(u-\tau)\biggl[\log^2\left(\frac{
c\,\tau}{2\,r}\right)+\frac{57}{70}\log\left(\frac{
c\,\tau}{2\,r}\right)+\frac{124627}{44100}\biggr]\nonumber \\
&-{428 \over 525}{G^3\, M^2 \over c^{11}}\,\log \left({r \over r_0}\right)\,I^{(3)}_{ij}(u)\,I^{(5)}_{ij}(u)\,,\label{Ftailoftailhar}\\
\mathcal{F}_\mathrm{(tail)^2}(u) &=
\frac{4\,G^3\,M^2}{5\, c^{11}}\left(\int_0^{+\infty}d\tau\,I^{(5)}_{
ij}(u-\tau)\biggl[\log\left(\frac{
c\,\tau}{2\,r}\right)+\frac{11}{12}\biggr]\right)^2\,.\label{Ftailsquaredhar}
\end{align}
\end{subequations}     
We shall come back to the discussion of  the hereditary terms in detail in Sec.~\ref{hereditary} where we shall compute their contributions to far-zone energy flux.
\section{The equations of motion and the conserved energy for Head-on collision}
\label{EqnOM}
\subsection{Standard Harmonic Coordinate System}
By standard harmonic coordinates we refer to the coordinate system that has been used in previous works~\cite{BFeom, Blanchet:2002mb}. 
Since the head-on collision problem has only one direction of motion one can write relevant equations for the head-on case by imposing the restrictions
\begin{eqnarray}
{\bf x}&=&z\,\hat{\bf n},\,\,\, {\bf v}=\dot{z}\,\hat{\bf n},\,\,\,r=z,\,\,\, v=\dot{r} = \dot{z}\,, 
\label{GOtoHOC}
\end{eqnarray}
on the corresponding expression for general orbits in terms of $z$ and $\dot {z}$, where $z$ is the separation between the two objects at a given time and $\dot {z}$ is the first time derivative of $z$, giving (coordinate) speed with which they are moving with respect to each other at that instant. 

The computation of the energy flux involves time derivatives of source multipole moments which in turn will require the knowledge of equations of motion at appropriate PN order. Computing the 3PN accurate energy flux requires the 3PN accurate equations of motion~\cite{Blanchet:2002mb}. 3PN accurate equations of motion in the CM frame associated with standard harmonic coordinate system, for compact objects moving in generic orbits, are given in~\cite{Blanchet:2002mb, ABIQ07}. 

Since we will discuss the results in other coordinate systems like the modified harmonic coordinates and the ADM coordinates in the subsequent sections, we will provide for a more general
discussion  of the 2.5PN terms along the lines of~\cite{Arun:2009mc} based on ~\cite{IW93,IW95} (see also~\cite{GII97}).
In addition to the contact transformation involving ``conservative'' orders up to 3PN required to go from the SH to the modified harmonic and ADM coordinates
(involving even order 2PN and 3PN terms) there still remains the possible change of gauge in the radiation reaction (dissipative) terms at order 2.5PN.  Recall, that in the SH 
coordinate system the lowest-order dissipative part of the equations of motion, i.e. the 2.5PN acceleration term, is given by (with boldface letters indicating ordinary three-dimensional vectors)
\begin{equation}\label{IWshar}
\mathbf{a}_{\rm 2.5PN}^{\rm SH} =\frac{8}{5}\frac{G^2 m^2\nu}{c^5 r^3}\left(\left[3\,v^2+\frac{17}{3}
\frac{G m}{r}\right] \,\dot{r}\,\mathbf{n} + \left[
-\,v^2 - 3\frac{G m}{r}\right]\,\mathbf{v} \right)\,.
\end{equation}
One may however  prefer to  employ alternative radiation gauges and a convenient characterization at 2.5PN order has been investigated  earlier in~\cite{IW93,IW95}. 
Following this work, the most general form of the 2.5PN term in the relative acceleration  is specified by the two-parameter family written as,
\begin{subequations}\label{IW}\begin{eqnarray}
\mathbf{a}_{\rm 2.5PN}^{\rm gen} &=& \frac{8}{5}\frac{G^2 m^2\nu}{c^5 r^3}\Bigl(A_{\rm 2.5PN}\,\dot{r}\,\mathbf{n}+B_{\rm 2.5PN}\,\mathbf{v}\Bigr)\,,\\
A_{\rm 2.5PN} &\equiv& 3(1+\beta)v^2+\frac{1}{3}\left(23+6\alpha-9\beta\right)\frac{G m}{r}-5\beta\dot{r}^2\,,\\
B_{\rm 2.5PN} &\equiv& -(2+\alpha)v^2-(2-\alpha)\frac{G m}{r}+3(1+\alpha)\dot{r}^2 \,.
\end{eqnarray}\end{subequations}
The general 2.5PN gauge is parametrized by the two numerical constants $\alpha$ and $\beta$. The SH (and modified harmonic) gauge in which the acceleration is given by \eqref{IWshar} 
corresponds to the choice $\alpha=-1$ and $\beta=0$; the ADM gauge corresponds to $\alpha=5/3$ and $\beta=3$, in which case the 2.5PN acceleration becomes~\cite{schafer82}
\begin{equation}\label{IWadm}
\mathbf{a}_{\rm 2.5PN}^{\rm ADM} =\frac{8}{5}\frac{G^2 m^2\nu}{c^5 r^3}\left(\left[12\,v^2+2\frac{G m}{r}-15\dot{r}^2\right] \,\dot{r}\,\mathbf{n} + \left[
-\frac{11}{3}\,v^2-\frac{1}{3}
\frac{G m}{r} + 8\dot{r}^2\right]\,\mathbf{v} \right)\,.
\end{equation}

By imposing the restrictions given by Eq.~\eqref{GOtoHOC} we can write the equations of motion (or acceleration) in terms of the variables $z$ and $\dot z$ for head-on situation as $a^i=a n^i$, where,
\begin{eqnarray}
\label{EOM}
a&=&-\frac{G\, m}{z^2}\Biggl\{1 + \frac{1}{c^2}\Bigg[\dot {z}^2\left(-3+\frac{7}{2}\,\nu\right)+\frac{G\, m}{z}\left(-4 -2\,\nu \right)\Bigg]\nonumber\\ 
&+&\frac{1}{c^4}\Bigg[\dot {z}^4\left(-\frac{21}{8}\, \nu-\frac{21}{8}\, \nu^2 \right)+\frac{G\, m}{z}\,\dot {z}^2\left(-11\,\nu+
4\, \nu^2\ \right)+\frac{ G^2\, m^2 }{z^2}\left(9+\frac{87}{4}\, \nu \right)
\Bigg]\nonumber\\  
&+&\frac{\dot z}{c^5}\Bigg[\frac{G\, m}{z}\,\dot z^2\left(-\frac{32}{5}-{16\over 5}\,\alpha+{16\over5}\,\beta\right)\nu+\frac{ G^2\, m^2}{z^2}\left(-\frac{136}{15}-{24\over 5}\,\alpha+{24\over5}\,\beta\right)\nu\Bigg]\nonumber\\
&+&\frac{1}{c^6}\Bigg[\dot {z}^6\left(-\frac{19}{16}\, \nu+\frac{13}{16}\, \nu^2+\frac{147}{16}\,\nu^3\right)\nonumber\\
&+&\frac{G\, m}{z}\,\dot {z}^4\left(\frac{199}{12}\, \nu-8\, \nu^2-12\, \nu^3\right)\nonumber\\
&+&\frac{G^2\, m^2}{z^2}\,\dot {z}^2\left(-3+\left[\frac{117709}{840}+\frac{123}{32}\,\pi^2-44\log \left(\frac{z}{z'_0}\right)\right]\nu+\frac{211}{8}\,\nu^2+2\,\nu^3 \right)\nonumber\\
&+&\frac{G^3\, m^3}{z^3}\left(-16+\left[-\frac{437}{4}+\frac{41}{16}\,\pi^2\right]\nu -\frac{71}{2}\,\nu^2\right)
\Bigg]
\Biggr\}+\mathrm{\mathcal{O}(7)}\,.
\end{eqnarray} 
The general expression for center-of-mass energy $E$ associated with standard harmonic coordinate system is given in~\cite{Blanchet:2002mb} and the corresponding expression for head-on situation can be obtained by imposing restrictions given by Eq.~\eqref{GOtoHOC}. 
Thus we have 
\begin{eqnarray}\label{ESH}
\frac{E_{\rm SH}(z,\dot{z})}{\mu}&=& -\frac{G\, m}{z}+\frac{\dot{z}^2}{2} \nonumber\\ &&
+\frac{1}{c^2}\bigg[\dot{z}^4\left(\frac{3}{8} - \frac{9}{8}\,\nu\right)+
\frac{G\, m}{z}\dot {z}^2\left(\frac{3}{2}+\nu \right)+\frac{1}{2}\frac{G^2\, m^2}{z^2}\bigg]\nonumber\\ &&
+\frac{1}{c^4}\bigg[\dot {z}^6\left(\frac{5}{16}-\frac{35}{16}\,\nu +
\frac{65}{16}\,\nu^2\right) 
+ \frac{G\, m}{z} \dot {z}^4\left(\frac{21}{8}
- {3\,\nu} - {6\,\nu^2} \right) 
+\frac{G^2\, m^2}{z^2} \dot {z}^2 \left
(\frac{9}{4} + \frac{7}{4}\,\nu+
{2\,\nu^2} \right) \nonumber\\
&&\qquad~
+\frac{G^3\, m^3}{z^3}\left( -\frac{1}{2}- \frac{15}{4}\,\nu
\right)\bigg]\nonumber\\ &&
+\frac{1}{c^6}\bigg[\dot {z}^8\left(\frac{35}{128} - \frac{413}{128}\,\nu +
\frac{833}{64}\,\nu^2 - \frac{2261}{128}\,\nu^3\right) \nonumber\\
&&\qquad~ + \frac{G\,m}{z}\dot {z}^6\left(\frac{55}{16} -{15\,\nu} +
\frac{25}{4}\,\nu^2 + {35\,\nu^3} 
\right) \nonumber\\ &&\qquad~ +
\frac{G^2\, m^2}{z^2} \dot {z}^4\left(\frac{147}{16} -\frac{569}{48}\,\nu -
\frac{245}{16}\,\nu^2 - 21\,\nu^3 \right)\nonumber\\ &&\qquad~ 
+\frac{G^3\, m^3}{z^3} \dot {z}^2\left( \frac{11}{4} +\left[-
\frac{9719}{420}- \frac{41}{32}\,\pi^2+\frac{44}{3}\log \Big(\frac{z}{z'_0}\Big)\right]\nu + \frac{15}{2}\,\nu^2 +{4\,\nu^3} 
\right)\nonumber\\ &&\qquad~ + \frac{G^4\, m^4}{z^4}\,\left(\frac{3}{8} +\left[\frac{18469}{840} -\frac{22}{3}\log\Big(\frac{z}{z'_0}\Big)\right]\nu\right)\bigg]+\mathrm{\mathcal{O}(8)}\,.
\end{eqnarray}
To study the head-on infall we consider two different situations, following~\cite{SPW95}. In the first case (we will call it case I) we assume that the radial infall proceeds from 
rest at a finite initial separation whereas in the second case (case II), we assume the objects start falling towards each other from  rest  at infinite separation.
\subsubsection{Case I: Infall from a finite distance} 
Let us suppose the two objects initially separated by the distance $z_i$ start falling radially towards each other from the rest $\it {i.e.}$ $\dot z\,(z_i)=0$. Hence the center-of-mass energy $E$ in standard harmonic coordinates at $z_i$ will be   
\begin{equation} 
E_{\rm SH}(z_{i},0) = -\mu\, c^2\,\gamma_{i} \left\{ 1 -{1 \over 2}\,\gamma_{i}
+\left[ {1\over 2}
+\frac{15}{4}\,\nu\right]\gamma_{i}^{2}+\left[-{3\over 8} + \left(-\frac{18469}{840}+ \frac{22}{3}\,\log{\left(\frac{z_{i}}{z'_{0}}\right)}\right)\nu\right]\,\gamma_{i}^{3}\right \}\,,
\label{En_zi_SH}
\end{equation}
where $\gamma_i= {G\,m/z_i c^2}$.
Equating, Eqs.~\eqref{ESH} and \eqref{En_zi_SH}, the resultant expression can be inverted  for ${\dot z}$. 
\begin{eqnarray}
\dot {z}(z,z_i)&=& -\sqrt {2}\,c\,{\sqrt {\gamma}\over \sqrt {1-s}}
\left \{1-s +\left[-{5\over2}+\frac{5}{4}\,\nu+s\left(3- {7\over2}\,\nu\right)+s^2\left(-{1\over2}+\frac{9}{4}\,\nu\right)\right]{\gamma} \right.\nonumber \\
& & \left.
+\left[{27\over8}-7\,\nu+ {55\over32}\, \nu^2 +s\left(-{37\over8} + \frac{179}{8}\,\nu-{173\over32}\, \nu^2\right)+s^2\left({13\over8}-\frac{119}{8}\,\nu+ {165\over32}\,\nu^2 \right)+s^3\left(-{3\over8}-\frac{1}{2}\,\nu-{47\over32}\, \nu^2 \right)\right]{\gamma^2}\right.\nonumber \\
& & \left.
+\left[-{65\over16}+\left[\frac{64343}{1120}+\frac{41}{32}\,\pi^2-11\log{\left(\frac{z}{z'_0}\right)}\right]\nu-\frac{945}{64}\,\nu^2+ {237\over128}\,\nu^3\right.\right.\nonumber \\
& & \left.\left.  
+s\left({23\over4}+\left[-\frac{73567}{560}-\frac{41}{32}\,\pi^2+\frac{44}{3}\log{\left(\frac{z}{z'_0}\right)}\right]\nu+\frac{1953}{32}\,\nu^2-{63\over8}\,\nu^3\right)\right.\right.\nonumber \\
& & \left.\left.  
+s^2\left(-{21\over8}+\frac{3271}{48}\,\nu-\frac{2507}{32}\,\nu^2+{633\over64}\,\nu^3\right)+s^3\left({5\over4}-{43\over16}\,\nu+\frac{729}{32}\,\nu^2-{65\over16}\,\nu^3\right)\right.\right.\nonumber \\
& & \left.\left.
+s^4\left(-{5\over16}+\left[\frac{28433}{3360}-\frac{11}{3}\log{\left(\frac{z_i}{z'_0}\right)}\right]\nu+\frac{595}{64}\, \nu^2+{25\over128}\,\nu^3\right)
\right]\gamma^3
\right\}\,,
\label{zdot-square-SH-FS}
\end{eqnarray}
where $\gamma= G\,m/z\,c^2$ is the PN parameter and $s=z/z_i< 1$.  
\subsubsection{Case II: Infall from infinity} 
We can view case II as a limiting case of case I and the expression for $\dot z$ can be obtained by inserting $s=z/z_i$ back in Eq.~\eqref{zdot-square-SH-FS} and taking the limit when $z_i \rightarrow \infty$. For $\dot {z}$ in SH coordinates, we have
\begin{eqnarray}
\dot {z}(z)&=&- {\sqrt{2}}\,c\,{\sqrt{\gamma}}
\left \{1+\left[-{5\over2}+\frac{5}{4}\,\nu
\right]{\gamma}
+\left[{27\over8}-7\,\nu+ {55\over32}\,\nu^2
\right]{\gamma^2}
\right.\nonumber \\
& & \left.
+\left[-{65\over16}+\left[\frac{64343}{1120}+\frac{41}{32}\,\pi^2-11\log{\left(\frac{z}{z'_0}\right)}\right]\nu-\frac{945}{64}\,\nu^2+ {237\over128}\,\nu^3
\right]\gamma^3
\right\}\,.
\label{zdot-square-SH-IS}
\end{eqnarray}
As expected, for $\nu=0$ the above relation is consistent with the radial geodesics  of Schwarzschild in \textit{standard harmonic coordinates}~\cite{SPW95} . 
\subsection{Modified Harmonic Coordinate System}
\label{EMH}
The SH coordinates are useful for analytical algebraic checks but also contain some gauge-dependent logarithms which are less suitable for numerical computations. It has been shown in~\cite{ABIQ07} that such dependences can be transformed away by using suitable gauge transformations. 
The expression for the shift ``$\delta_{\rm (SH\rightarrow MH)}E$'' for general orbit case has been given by Eq.~(4.12) of~\cite{ABIQ07} and we have used Eq.~\eqref{GOtoHOC} to obtain the corresponding expression for head-on situation.
We can write the center-of-mass energy in MH coordinates using the relation
\begin{equation}
\label{ruleESHtoMH}
E_{\rm MH}=E_{\rm SH}+\delta_{(\rm SH \rightarrow MH)}E\,,
\end{equation}
where $E_{\rm SH}$ represents the energy $E$ in SH coordinates and is given by Eq.~\eqref{ESH}. ``$\delta_{\rm (MH \rightarrow SH)}E$'' for head-on case reads 
\begin{equation}\label{ESHtoMH}
\delta_{(\rm SH\rightarrow MH)}E = \frac{22}{3}\frac{G^3\,m^4\,\nu^2}{c^6\,z^3}\left\{
\biggl[\frac{G\,m}{z}-2\,\dot{z}^2 \biggr]\log\left(\frac{z}{z_0'}\right)
+\dot{z}^2\right\} + \mathcal{O}(8)\,.
\end{equation}
\subsubsection{Case I: Infall from a finite distance} 
It is evident from the above that using Eqs.~\eqref{ESH} and \eqref{ESHtoMH} in Eq.~\eqref{ruleESHtoMH} we can write the expression for conserved energy $E$ in MH coordinates. At the initial separation $z_i$ energy in MH coordinates reads
\begin{equation}
E_{\rm MH}(z = z_{i}) = -\mu c^2 \gamma_{i} \left\{ 1 -{1 \over 2}\,\gamma_{i}
+\left[ {1 \over 2}
+\frac{15}{4}\,\nu\right] \gamma_{i}^{2}+\left[-{3\over 8} - \frac{18469}{840}\,\nu\right]\gamma_{i}^{3}\right\}\,.
\label{En_zi_MH}
\end{equation}
By equating Eq.~\eqref{En_zi_MH} and the expression for center-of-mass energy in MH coordinates and then inverting the resultant expression, one can obtain the expression for $\dot z$ in MH coordinates. For brevity in presentation, in what follows, we will list only the differences in various expressions in a particular coordinate system from their SH values. By adding the difference to the SH expression the corresponding expression in the relevant coordinate can be computed. In particular in this case, for ${\dot z}$ we have,
\begin{equation}
\label{rulezdot2SHtoMH}
{\dot{z}}_{\rm MH}={\dot z}_{\rm SH}+\delta_{\rm (SH\rightarrow MH)}{\dot z},
\end{equation}
where,
\begin{eqnarray}
\delta_{\rm (SH \rightarrow MH)}{\dot z}&=&-\sqrt{2}\,c\,{\gamma^{7/2}\over \sqrt{1-s}}\left\{
\left(-{22 \over 3}+11\log\left({z\over z_0'}\right)\right)\nu
+s\left({22 \over 3}-{44\over 3}\log \left({z\over z_0'}\right)\right)\nu\right.\nonumber\\
& & \left.
+s^4\left({11\over 3}\log \left({z_i\over z_0'}\right)
\right)\nu
\right\}.
\label{zdot-square-MH-FS}
\end{eqnarray}
Though to avoid heavy notation we write $z$ and $\dot{z}$, beware that
in this subsection they correspond to $z_{\rm MH}$ and $\dot{z}_{\rm MH}$ respectively
and in the next subsection to $z_{\rm ADM}$ and $\dot{z}_{\rm ADM}$ respectively.
\subsubsection{Case II: Infall from infinity} 
Once again, by inserting $s=z/z_i$ back in Eq.~\eqref{zdot-square-MH-FS} and taking the limit when $z_i \rightarrow \infty$ we obtain the expression for ``$\delta_{\rm (SH \rightarrow MH)}{\dot {z}}$'' as
\begin{equation}
\delta_{\rm (SH \rightarrow MH)}{\dot z}=-\sqrt {2}\,c\,\gamma^{7/2}
\left[
-{22 \over 3}+11\log\left({z\over z_0'}\right)
\right]\nu\,.
\label{zdot-square-MH-IS}
\end{equation}
\subsection{ADM coordinate System}
\label{EADM}
Finally, in this section we provide the expressions for the conserved energy in ADM coordinates. Like MH coordinate systems the ADM coordinate system is also free from logarithms appearing in 3PN expressions of EOM or source multipole moments when standard harmonic coordinate system is used. We can write the center-of-mass energy in ADM coordinates using the relation 
\begin{equation}
E_{\rm ADM}=E_{\rm SH}+\delta_{\rm (SH\rightarrow ADM)}{E}, 
\end{equation}
where $E_{\rm SH}$ is given by Eq.~\eqref{ESH} and for ``$\delta_{\rm (SH\rightarrow ADM)}{E}$'' in head-on situation (see Appendix~\ref{deltaESHtoADM} for its computation) we have 
\begin{align} \label{EnSHtoADM}
\delta_{\rm (SH\rightarrow ADM)}{E} =& \frac{G^2\, m^3\, \nu}{c^4\, z^2}\left\{\dot {z}^2\left(-\frac{1}{4}-\frac{5}{4}\,\nu\right)+\frac{G\,m}{z}\left({1 \over 4}+3\,\nu \right)\right\}
\nonumber\\&+\frac{G^2\, m^3 \,\nu }{c^{6}\, z^2}
\left\{\dot z^4 \left(-{3 \over 8}+\frac{20}{3}\,\nu+4\,\nu^2\right)
 \right.\nonumber\\&\qquad +\frac{G\,m}{z}\dot z^2 \left(-\frac{9}{8} +\left[\frac{5441}{280}+\frac{21}{16}\,\pi^2-\frac{44}{3}
\,\log\left(\frac{z}{z'_0}\right)\right]\nu-\frac{55}{8}\,\nu^2\right)
\nonumber\\&\qquad \left.
+\frac{G^2\, m^2}{z^2}\left(-{1 \over 4} +\left[-\frac{3613}{280} -\frac{21}{32}\,\pi^2+\frac{22}{3}\log \left(\frac{z}{z'_0}\right)\right]\nu\right)\right\}\,.
\end{align}  
\subsubsection{Case I: Infall from a finite distance} 
\begin{equation}
E_{\rm ADM}(z = z_{i}) = -\mu\, c^2\, \gamma_{i} \left\{ 1 -{1 \over 2} \gamma_{i}
+\left[\frac{1}{4}+ \frac{3}{4}\,\nu\right]\gamma_{i}^{2}+\left[-{1\over 8} +\left(- \frac{109}{12}+\frac{21}{32}\,\pi^2\right)\nu\right]\gamma_{i}^{3}\right \}\,.
\label{En_zi_ADM}
\end{equation}
For ${\dot z}$ in ADM coordinate we have,
\begin{equation}
\label{rulezdot2SHtoADM}
{\dot{z}}_{\rm ADM}={\dot z}_{\rm SH}+\delta_{\rm (SH\rightarrow ADM)}{\dot z}\,,
\end{equation}
where,
\begin{eqnarray}
\delta_{\rm (SH \rightarrow ADM)}{\dot z}&=&-\sqrt{2}\,c\,{\sqrt{\gamma}\over \sqrt{1-s}}\left\{
\left[{1\over 8}-{\nu \over 4}+s\left(-{1\over4}-{5\over4}\,\nu\right)+s^3\left({1\over8}+{3\over2}\,\nu\right)
\right]\gamma^2\right.\nonumber \\
& & \left.
+\left[{5\over16}+\left(-{101959 \over 3360}-{63\over64}\,\pi^2+11\log\left({z\over z_0'}\right)\right)\nu+{9\over16}\,\nu^2\right.\right.\nonumber 
\\ & & \left.\left.
+s\left({1 \over 16}+\left[{179147\over3360}+{21\over16}\,\pi^2-{44\over 3}\log \left({z\over z_0'}\right)\right]\nu-{41\over8}\,\nu^2\right)\right.\right.\nonumber 
\\ & & \left.\left.
+s^2\left(-{1\over8}-{715\over48}\,\nu+{97\over16}\,\nu^2\right)
+s^3\left(-{7\over16}-{145\over32}\,\nu+{69\over8}\,\nu^2\right)\right.\right.\nonumber 
\\ & & \left.\left.
+s^4\left({3\over16}+\left[-{3971\over1120}-{21\over64}\,\pi^2+{11\over 3}\log \left({z_i\over z_0'}\right)\right]\,\nu-{81\over8}\,\nu^2\right)\right]\gamma^{3}
\right\}\,.
\label{zdot-square-ADM-FS}
\end{eqnarray}
\subsubsection{Case II: Infall from infinity} 
The expression for ``$\delta_{\rm (SH \rightarrow ADM)}{\dot{z}}$'' can be written by inserting $s=z/z_i$ back in Eq.~\eqref{zdot-square-ADM-FS} and taking the limit when $z_i \rightarrow \infty$ as
\begin{eqnarray}
\delta_{\rm (SH \rightarrow ADM)}{\dot z}&=&\sqrt{2}\,c\,\sqrt{\gamma}\left\{
\left({1\over 8}-{\nu \over 4}
\right)\gamma^2
\right.\nonumber \\
& & \left.
+\left(\left[{5\over16}+\left(-{101959 \over 3360}-{63\over64}\,\pi^2+11\log\left({z\over z_0'}\right)\right)\nu+{9\over16}\,\nu^2\right]
\right)\gamma^{3}
\right\}\,.
\label{zdot-square-ADM-IS}
\end{eqnarray}
\subsection{Inputs for the computation of the hereditary part}
\label{inputs-hered}
It is evident from Eqs.~\eqref{Ftailhar}-\eqref{Ftailtailhar} that all integrals need to be evaluated with just Newtonian order accuracy except the one in the first term of Eq.~\eqref{Ftailhar} which needs to be computed with 1PN accuracy and hence in this section we provide 1PN accurate inputs which will be required for the computation of hereditary part of the energy flux. Since at 1PN order the expressions for all desired inputs [{\it {e.g.}} source moments, trajectory of the problem and relation connecting ADM mass to total mass ($m=m_1+m_2$)] are the same in all the three coordinate systems we need not give these inputs in different coordinate systems.\\
\subsubsection{ Case I: Infall from a finite distance} 
Equation~\eqref{zdot-square-SH-FS} gives the expression for $\dot z$ which in the 1PN limit can be expressed as
\begin{equation}
\dot {z}=-\sqrt{{2\,G\,m\over z}}\,\sqrt{1-s}\,\left[1+\,{G\,m\over c^2\,z}\left(-{5\over2}+{5\over4}\,\nu+s\left({1\over2}-{9\over4}\,\nu\right)\right)\right]\,.
\label{zdot-FS}
\end{equation}
Solving Eq.~\eqref{zdot-FS} we get the  1PN trajectory as 
\begin{equation}
t={z^{3/2}_i\over \sqrt {2\,G\,m}}\left[g(s)-
{1\over 2}{G \,m \over c^2 \,z_i} \,h_0(s)
-\nu\,{1\over 2}{G \,m \over c^2 \,z_i}
 h_1(s)\right]\,,
\label{trajectory_FS}
\end{equation}
where $g(s)$, $h_0(s)$ and $h_1(s)$ are the following 
simple linear combinations
of the elementary functions $f_1(s)=\sqrt{s}\sqrt{1-s}$ and
$f_2(s)=\arcsin{\sqrt{s}}$,
\begin{subequations}
\label{gs-hs}
\begin{eqnarray}
g(s)&=&f_1(s)-f_2(s)\,,\\
h_0(s)&=&f_1(s)+9f_2(s)\,,\\
h_1(s)&=&-\frac{1}{2}(9f_1(s)+f_2(s))\,.
\end{eqnarray}
\end{subequations}
\subsubsection{ Case II: Infall from infinity} 
In the 1PN limit Eq.~\eqref{zdot-square-SH-IS} gives
\begin{equation}
\dot {z}=-\sqrt{{2\,G\,m\over z}}\,\left[1-{5\over2}\,{G\,m\over c^2\,z}\left(1-{\nu \over2}\right)\right]\,.
\label{zdot-IS}
\end{equation}
By integrating Eq.~\eqref{zdot-IS} we get the 1PN trajectory of the problem which reads 
\begin{equation}
t = -\frac{\sqrt{2}\, z^{3/2}}{3\, \sqrt{G}\, \sqrt{m}}\left[1+{15 \over 2}\frac{G\, m}{c^2\, z}\left(1-\frac{\nu}{2}\right)\right]\,.
\label{trajectory-IS}
\end{equation}
\section{The multipole moments of compact binary systems}
\label{sourcemoments}
In this section we shall provide the 
expressions for source multipole moments with an accuracy sufficient for the computation of the 3PN accurate energy flux in standard harmonic coordinates. General expressions for these  moments have been given in~\cite {ABIQ07} for inspiralling compact objects moving in generic orbits in standard harmonic coordinates. 
Since the head-on collision problem has only one direction of motion one can write the expressions for source moments for head-on case by imposing the restrictions,
Eq.~\eqref{GOtoHOC} on the corresponding expression for general orbits in terms of $z$ and $\dot {z}$.
Further, as discussed in~\cite{Arun:2009mc}, the 2.5PN term for general orbits in SH coordinates (see Eq. (3.1) of \cite {ABIQ07}) is modified by
\begin{equation}
\label{IijSHtogen}
\delta_{\epsilon}I_{ij}=-{16 \over 15}{G^2\,m^3\,\nu^2\over c^5}\left[-\beta\,{\dot {r}}\,n_{\langle i}n_{j \rangle}+\left(3+3\,\alpha-2\,\beta\right)n_{\langle i}v_{j \rangle}\right]\,, 
\end{equation}
which for the head-on case reduces to $-{16\over5}{G^2\,m^3\,\nu^2\over c^5}\,{\dot {z}}\left(1+\alpha-\beta\right)\,n_{\langle i}n_{j \rangle}$.
Thus, the 3PN mass quadrupole $I_{ij}$  for head-on case reads as
\begin{eqnarray}
 I_{ij} &=& m\, z^2\, \nu\,\left\{1+\frac{1}{c^2}\left[\dot {z}^2\left(\frac{9}{14} - \frac{27}{14}\, \nu\right) + \frac{G\,m}{z}\left(-\frac{5}{7} + \frac{8}{7}\, \nu\right)
\right]\right.\nonumber \\ & & \left.
+\frac{1}{c^4}\Biggl[\dot {z}^4\left(\frac{83}{168} - \frac{589}{168}\, \nu+\frac{1111}{168}\, \nu^2\right)\right.\nonumber \\ & & \left.
+\frac{G\, m}{z}\dot {z}^2\left(\frac{32}{9} + \frac{289}{126}\, \nu - \frac{1195}{126}\,\nu^2\right)\right.\nonumber \\ & & \left.
+\frac{G^2\, m^2}{z^2}\left(-\frac{355}{252} - \frac{953}{126}\, \nu + \frac{337}{252}\, \nu^2\right)\Biggr]\right.\nonumber \\ & & \left.
+\frac{\dot z}{c^5}\left[\frac{G^2\, m^2}{z^2}\left({8\over35}-{16\over5}\alpha+{16\over5}\beta\right)\nu\right]\right.\nonumber \\ & & \left.
+\frac{1}{c^6}\Biggl[\dot {z}^6\left(\frac{507}{1232} - \frac{6101}{1232}\, \nu + \frac{12525}{616}\, \nu^2 - \frac{34525}{1232}\, \nu^3\right)\right.\nonumber \\ & & \left.
+\frac{G\, m}{z}\dot {z}^4 \left(\frac{26177}{5544} - \frac{9313}{693}\, \nu - \frac{34165}{2772}\, \nu^2 + \frac{75787}{1386}\, \nu^3\right)\right.\nonumber \\ & & \left.
+\frac{G^2\,m^2}{z^2}\dot {z}^2\left(-\frac{8281261}{415800}+\frac{428}{105}\log\left(\frac{z}{z_0}\right)+\frac{15703}{792}\,\nu + \frac{13803}{616}\, \nu^2 - \frac{527797}{16632}\, \nu^3\right)\right.\nonumber \\ & & \left.
+\frac{G^3\,m^3}{z^3}\left(\frac{6285233}{207900}- \frac{428}{105}\log
\left(\frac{z}{z_0}\right) + \left[\frac{15502}{385}-\frac{44}{3}\log\left(\frac{z}{z_0'}\right)\right]\nu - \frac{3632}{693}\, \nu^2\right.\right.\nonumber \\ & & \left.\left.+\frac{13289}{8316}\,\nu^3\right)
\Biggr]
\right\}n_{\langle i}n_{j \rangle}+\mathrm{\mathcal{O}(7)}\,.\label{MQHOC}
\end{eqnarray} 
Note that the quantity $z_0$ appearing in above expression is, in our
present head-on notation, the 
constant length scale $r_0$ appearing in Eq.~\eqref{U-u} and in the relations connecting radiative multipole moments and source multipole moments. The presence of the other constant $z_0'$ through some logarithms $\log (z/z_0')$ at 3PN order is due to the use of standard harmonic coordinates and corresponds to 
$r_0'$ in the earlier papers. Later in this paper we shall show that alternatively one can use other coordinate systems such as
the  MH or ADM coordinate system which do not involve such logarithms.\\
The 2PN mass octupole $I_{ijk}$ for head-on case is given by
\begin{eqnarray}
I_{ijk}&=& -m\, z^3 \,\sqrt{1-4\, \nu }\, \nu \Biggl\{1+\frac{1}{c^2}\Big[\dot {z}^2\left(\frac{5}{6}-\frac{19}{6}\,\nu\right)+\frac{G\, m}{z} \left(-\frac{5}{6}+\frac{13}{6}\,\nu\right)\Big] \nonumber \\
&+&\frac{1}{c^4}\Big[\dot {z}^4\left(\frac{61}{88}-\frac{1579}{264}\,\nu+\frac{1129}{88}\,\nu ^2\right) \nonumber \\
  &+&\frac{G\, m}{z}\dot {z}^2 \left(\frac{54}{11}+\frac{521}{132}\,\nu-\frac{2467}{132}\,\nu^2 \right) \nonumber \\ 
&+&\frac{G^2\, m^2}{z^2} \left(-\frac{47}{33}-\frac{1591}{132}\,\nu+\frac{235}{66}\,\nu ^2\right)
\Big]
\Biggr\} n_{\langle i}n_{j }n_{k \rangle}+\mathrm{\mathcal{O}(5)}\,.\label{MOHOC}
\end{eqnarray}
The 1PN mass moment, $I_{ijkl}$  reads as
\begin{eqnarray}
I_{ijkl}&=& m\,z^4\, \nu\, \Biggl\{1-3\, \nu+\frac{1}{c^2}\Big[\dot {z}^2\left(\frac{23}{22}-\frac{159}{22}\,\nu+\frac{291}{22}\, \nu ^2\right) \nonumber \\
&+&\frac{G\, m}{z} \left(-\frac{10}{11}+\frac{61}{11}\,\nu -\frac{105}{11}\,\nu ^2\right)\Big]
\Biggr\}n_{\langle i}n_{j }n_{k}n_{l \rangle}+\mathrm{\mathcal{O}(4)}\, .\label{M4HOC}
\end{eqnarray}
The moment, $I_{ijklm}$ which will be needed with Newtonian accuracy is
\begin{eqnarray}
I_{ijklm}&=&-m\, z^5\, \sqrt{1-4\, \nu }\, (1-2\, \nu )\, \nu\, n_{\langle i}n_{j }n_{k}n_{l}n_{m \rangle}+\mathrm{\mathcal{O}(2)}\,.\label{M5HOC}
\end{eqnarray}
Finally we give the monopole moment $W$, which is
\begin{equation}
W={1 \over 3}\,\nu\,m\,z\,\dot {z}+\mathrm{\mathcal{O}(2)}\,.
\label{W}
\end{equation} 

\section{Instantaneous contributions in the Energy Flux for head-on infall}
\label{3PNEFINST}
\subsection{The 3PN instantaneous energy flux in Standard Harmonic Coordinates}
\label{SH}
Having source multipole moments given by Eqs.~\eqref{MQHOC}-\eqref{W} and equations of motion given by Eq.~\eqref{EOM} with the desired PN accuracies one can compute the  required time derivatives of source moments to get instantaneous contribution to the  far-zone GW energy flux using Eq.~\eqref{inst}. Since the instantaneous contribution to 3PN far-zone energy flux for compact binaries moving in general orbits has already been listed in~\cite{ABIQ07} in terms of SH variables $r$, $\dot r$ and $v$, we can also directly write down the corresponding expression for instantaneous energy flux in terms of the variables $z$ and $\dot {z}$ for the head-on situation using Eq.~\eqref{GOtoHOC}.  The form of the 2.5PN terms in the flux for a general 2.5PN gauge is discussed in~\cite{Arun:2009mc} [see Eq.(3.14a) there],
and we adapt it for the head-on case. We write the  result as,   
\begin{equation}
\label{EF} \mathcal{F}_\mathrm{inst} = \mathcal{F}^\mathrm{N}_\mathrm{inst}+
\mathcal{F}^\mathrm{1PN}_\mathrm{inst}+ \mathcal{F}^{\rm
2PN}_\mathrm{inst}+\mathcal{F}^\mathrm{2.5PN}_\mathrm{inst}+
\mathcal{F}^\mathrm{3PN}_\mathrm{inst} +\mathcal{O}(7)\,,
\end{equation}
and find that the various PN pieces for the head-on case are given by
\begin{subequations} \label{3PNEF-inst}
\begin{align}
\mathcal{F}^\mathrm{N}_\mathrm{inst}&={\frac{8}{15} \frac{G^3\,
m^4\,\nu^2}{c^5\, z^4}\, \dot{z}^2}\,,
\label{0PNEF}\\
\mathcal{F}^\mathrm{1PN}_\mathrm{inst}&={\frac{8}{15}\frac{G^3\,
m^4\,\nu^2}{c^7\, z^4}\left\{\dot {z}^4\left(-\frac{32}{7} + \frac{18}{7}\,\nu\right) 
+\frac{G\, m}{z}\dot {z}^2 \left(\frac{54}{7}+\frac{10}{7}\,\nu \right)
+\frac{G^2\, m^2}{z^2} \left(\frac{4}{7}-\frac{16}{7}\,\nu\right)\right\}}\,,
\label{1PNEF}\\
\mathcal{F}^\mathrm{2PN}_\mathrm{inst}&={\frac{8}{15}\frac{G^3\,
m^4\,\nu^2}{c^9\, z^4}\left\{\dot {z}^6\left(\frac{44}{21}-\frac{25}{21}\,\nu
 -\frac{67}{21}\,\nu^2\right)\right.}
{+\frac{G\, m}{z}
\dot {z}^4\left(-\frac{655}{21}+\frac{65}{63}\,\nu +\frac{1016}{63}\,\nu^2\right)}\nonumber\\& 
{+\frac{G^2\, m^2}{z^2}
\dot {z}^2\left(-\frac{9371}{189}+\frac{971}{14}\,\nu -\frac{140}{9}\,\nu^2 \right)}
{\left.+\frac{G^3\, m^3}{z^3}
\left(-\frac{506}{63}+\frac{228}{7}\,\nu-\frac{16}{9}\,\nu^2\right)\right\}}\,,
\label{2PNEF}\\
\mathcal{F}^\mathrm{2.5PN}_\mathrm{inst}&=
\frac{8}{15}\frac{G^3\, m^4\, \nu^2}{c^{10}\,z^4}\,\dot {z}
\left\{\frac{8}{7}\frac{G\, m}{z}\dot{z}^4 
+\frac{G^2\, m^2}{z^2}\dot {z}^2\left({96\over7}+{48\over5}\,\alpha-{48\over5}\,\beta\right) +
\frac{G^3\, m^3}{z^3}\left({332\over105}+{16\over5}\,\alpha-{16\over5}\,\beta\right)\right\}\nu\,,
\label{2p5PNEF}\\
\mathcal{F}^\mathrm{3PN}_\mathrm{inst}&={\frac{8}{15}\frac{G^3
\,m^4\,\nu^2}{c^{11}\, z^4}\left\{
\dot {z}^8\left(\frac{360}{77}-\frac{3533}{924}\,\nu
-\frac{247}{462}\,\nu^2+\frac{806}{77}\,\nu^3\right)\right.}\nonumber\\&
{+\frac{G \,m}{z}
\dot {z}^6\left(-\frac{863}{165}+\frac{24842}{693}\,\nu
-\frac{33515}{462}\,\nu^2-\frac{65705}{2772}\,\nu^3\right)}
\nonumber\\& 
{+\frac{G^2
\,m^2}{z^2}\dot {z}^4\left(-\frac{840109}{9450}+\frac{856}{35}
\log\left(\frac{z}{z_0}\right)\right.}\nonumber\\&
{\left.-\left[\frac{234205}{396}+\frac{123}{16}\,\pi^2\right] \nu
+\frac{418973}{1386}\,\nu^2+\frac{176927}{4158}\,\nu^3\right)}\nonumber\\&
{+\frac{G^3\, m^3}{z^3}
\dot {z}^2\left(-\frac{11963741}{51975}+\frac{3424}{105}
\log\left(\frac{z}{z_0}\right)\right.}\nonumber\\&
{\left.-\left[\frac{22682}{315}+\frac{205}{8}\,\pi^2-
\frac{176}{3}
\log\left(\frac{z}{z'_0}\right)\right]\nu
+\frac{72477}{154}\,\nu^2 -\frac{20147}{231}\,\nu^3\right)}\nonumber\\&
{\left.+\frac{G^4\, m^4}{z^4}
\left(\frac{37571}{693}-\frac{59848}{297}\,\nu-\frac{6038}{99}\,\nu^2-\frac{3464}{2079}\,\nu^3\right)\right\}}\,. \label{3PNEF}
\end{align}
\end{subequations}

The dependence of the result \eqref{EF}--\eqref{3PNEF-inst} on $z_0'$ is 
due to our use of the SH coordinate system. We will transform away this dependence by making use of a different coordinate system such as MH coordinate system. The presence of constant $z_0$ is not surprising as it was present in the expression of the mass quadrupole moment and hence appears in the final expression for the instantaneous part of the 3PN energy flux. This dependence of the instantaneous terms on the constant $z_0$ should exactly cancel a similar contribution coming from the tail terms. We explicitly show this cancellation in the next section.

The general expression for the energy flux above for the
head-on case takes a simpler form for radial infall from rest.
In this case the velocity $\dot{z}$ can be expressed solely in terms of the 
coordinate $z$ and the initial coordinate separation 
where it is at rest as shown in Sec.~\ref{EqnOM}.
However, since we are working up to 3PN there is one last element to
be taken into account before we can proceed. This relates to the 
infall velocity due to leading gravitational radiation reaction
that induces a $\dot{z}_{\rm RR}$ at 2.5PN i.e. $\dot{z}_{\rm 2.5PN}$.
To evaluate this we can adapt the treatment in~\cite{KBI07} to the head-on
case. It requires only the leading term in $E$ and the GW energy flux ${\cal F}$ and the infall due to  radiation reaction for the finite separation
case is given by
\begin{equation} 
\dot{z}_{\rm RR}=\frac{dE/dt}{dE/dz}=-\frac{{\cal F}}{dE/dz}=
-\frac{16}{15}\frac{G^3 m^3}{c^5 z^3} \nu (1-s)\,.
\label{zdotRR}
\end{equation} 
Adding $\dot{z}_{\rm RR}$ to the $\dot{z}$ given by 
Eq.~\eqref{zdot-square-SH-FS}
yields the complete 3PN accurate $\dot{z}$ that will be employed
in the next subsection to rewrite the energy flux solely in terms
of the variable $z$.
It should be obvious that the form of $\dot{z}_{\rm RR}$ is the same
in all the three coordinate systems that we use in this paper.

We now have all basic ingredients for the computation of the 3PN GW energy flux from compact objects with arbitrary mass ratios falling radially towards each other and can proceed to compute the instantaneous part of the energy flux for the head-on infall case.\\ 
\subsubsection{Case I: Infall from a finite distance} 

Starting from the 3PN instantaneous contribution to  energy flux in SH coordinates [Eq.~\eqref{EF}-\eqref{3PNEF-inst}] in terms of the variables $z$ and $\dot{z}$ and substituting the expression for $\dot z$ given by Eq.~\eqref{zdot-square-SH-FS} supplemented by $\dot{z}_{\rm RR}$ given by
Eq.~\eqref{zdotRR} we get the final expression for energy flux in terms of just one variable $z$. The final expression for the instantaneous part of energy flux in standard harmonic coordinate reads as
\begin{eqnarray}
\left( {d{\cal E} \over dt}\right)_{\rm SH}^{\rm inst} 
&=& { 16 \over 15}\,\frac{c^5}{G}\,\nu^2\,\gamma^5
\left \{ 1 - s +\left [-{43\over 7} +{111\over 14}\, \nu 
+s\left( {116\over 7} -{131\over 7}\,\nu \right ) + s^2\left( -{71\over 7} +{135\over 14}\,\nu \right)
\right ]{\gamma}\right.\nonumber \\
& & \left.
+ \left [ -{ 1127 \over 27} - {803 \over 36}\,\nu
+{112 \over 3}\, \nu^2
+ s \left(- { 4471 \over 189} + {15481\over 63}\,\nu 
- {2864\over 21}\,\nu^2 \right )\right.\right.\nonumber \\
& & \left.\left.
+ s^2 \left ( {1870 \over 21}
-{ 5503\over 18}\,\nu + { 8800\over 63}\,\nu^2\right )
+s^3 \left ( -{83 \over 3} +{1183\over 12}\,\nu  - {872\over 21}\, \nu^2\right )
\right]
{\gamma^2}\right.\nonumber \\
& & \left.
+\sqrt {2}\sqrt {1-s}\left[\left(-\frac{578}{35}-{56\over5}\,\alpha+{56\over5}\,\beta\right)\nu+s\left({1808\over105}+{48\over5}\,\alpha-{48\over5}\,\beta\right)\nu-\frac{16}{7}\,s^2\,\nu \right]{\gamma^{5/2}}\right.\nonumber \\ & & \left. 
+\left[-\frac{195203}{20790}+
\frac{1712}{21}\log{\left(\frac{z}{z_{0}}\right)}+\left[-\frac{2855087}{2772}-\frac{615}{16}\,\pi^2+\frac{110}{3}\log{\left(\frac{z}{z'_{0}}\right)}\right]\nu+\frac{1738129}{5544}\,\nu^2+\frac{231520}{2079}\,\nu^3
\right.\right.\nonumber \\
& & \left.\left. + s\left(\frac{4260593}{17325}-\frac{13696}{105}\log{\left(\frac{z}{z_{0}}\right)}+
\left[\frac{23954477}{41580}+\frac{861}{16}\,\pi^2-\frac{88}{3}\log{\left(\frac{z}{z'_{0}}\right)}\right]\nu+
\frac{3228377}{2772}\,\nu^2-\frac{1366537}{2079}\,\nu^3 \right)\right.\right.\nonumber \\
& & \left.\left.+ s^2\left(-\frac{19591864}{51975}+\frac{1712}{35}\log{\left(\frac{z}{z_{0}}\right)}+\left[\frac{523423}{308}-\frac{123}{8}\,\pi^2\right]\nu-\frac{47223}{14}\,\nu^2+\frac{304961}{297}\,\nu^3\right)\right.\right.\nonumber \\
& & \left.\left.
+s^3\left(\frac{212164}{1155}-\frac{2192423}{1386}\,\nu+\frac{897295}{396}\,\nu^2-\frac{406498}{693}\,\nu^3\right)\right.\right.\nonumber \\
& & \left.\left.
+s^4\left(-\frac{1245}{77}+\left[\frac{1088117}{4620}-\frac{22}{3}\log{\left(\frac{z_{i}}{z'_{0}}\right)}\right]\nu-\frac{105989}{264}\,\nu^2+\frac{8076}{77}\,\nu^3\right)\right]{\gamma^3} \right \}\,.
\label{EFSH-FS}
\end{eqnarray}
The standard harmonic gauge at 2.5PN  corresponds to $\alpha=-1$ and $\beta=0$.
\subsubsection{ Case II: Infall from infinity} 
As discussed in the previous section, we can see case II as a special case of case I and the expression $ { d{\cal E}/ dt } $ can be obtained by inserting $s=z/z_i $ in Eq.~\eqref{EFSH-FS} and taking the limit when $z_i \rightarrow \infty$ . The instantaneous part of energy flux in SH coordinates reads,
\begin{eqnarray}
\left( {d{\cal E} \over dt}\right)_{\rm SH}^{\rm inst} 
&=& { 16 \over 15}\,\frac{c^5}{G}\,\nu^2\,\gamma^5
\left \{ 1  +\left [- \frac{43}{7} +{111 \over 14}\,\nu
\right ]{\gamma}
+ \left [- { 1127 \over 27} - {803 \over 36}\,\nu
+{112 \over 3}\, \nu^2
\right ]
{\gamma^2}
+\sqrt {2}\left[\left(-\frac{578}{35}-{56\over5}\,\alpha+{56\over5}\,\beta\right)\nu\right]{\gamma^{5/2}}\right.\nonumber \\ & & \left. 
+\left[-\frac{195203}{20790}+\frac{1712}{21}\log{\left(\frac{z}{z_{0}}\right)}+\left[-\frac{2855087}{2772}-\frac{615}{16}\,\pi^2+\frac{110}{3}\log{\left(\frac{z}{z'_{0}}\right)}\right]\nu+\frac{1738129}{5544}\,\nu^2\right.\right.\nonumber \\& & \left.\left.
+\frac{231520}{2079}\,\nu^3\right]{\gamma^3} \right \}\,.
\label{EFSH-IS}
\end{eqnarray}
The standard harmonic gauge at 2.5PN  corresponds to $\alpha=-1$ and $\beta=0$.
\subsection{The 3PN instantaneous Energy Flux in Modified Harmonic Coordinates}
\label{MH}
As we have pointed out in Sec.~\ref{EMH} that one needs to use an alternative coordinate system such as MH coordinate system, 
which  is more suitable for numerical computations. 
In this section we provide
the 3PN energy flux expressions in MH coordinates. We can write the energy flux $\mathcal{F}$ in the MH coordinates by using the relation 
\begin{equation}
\mathcal{F}_{\rm MH}=\mathcal{F}_{\rm SH}+\delta_{\rm (SH\rightarrow MH)}\mathcal{F}\,,
\end{equation}
where $\mathcal{F}_{\rm SH}$ is the energy flux in SH coordinates for head-on situation  given by Eqs.~\eqref{EF}-\eqref{3PNEF-inst}. The general expression for the shift ``$\delta_{\rm (SH\rightarrow MH)}\mathcal{F}$'' is given by Eq.~(6.8) of ~\cite{ABIQ07}, which in the head-on situation reduces to the following form when we use the restrictions given by Eq.~\eqref{GOtoHOC},    
\begin{equation} \label{EFSHtoMH}
\delta_{\rm (SH\rightarrow MH)}\mathcal{F} =- \frac{1408}{15}\,\frac{G^6 \,m^7
\,\nu^3}{c^{11}\,z^7}\left[ \frac{1}{3} \dot z^2 \log\left(\frac{z}{z'_0}\right) - \frac{\dot{z}^2}{12} +
\mathcal{O}(2)\right]\,.
\end{equation}
\\  
Once we have expressions for the energy flux $\mathcal{F}$ in MH coordinates, we can compute the final expression for instantaneous part of energy flux in MH coordinates following the procedure adopted in Sec.~\ref{SH}.
\subsubsection{Case I: Infall from a finite distance} 
Substituting for $\dot {z}$ in MH coordinates, in the expression for energy flux in MH coordinates, one can obtain the expression for the instantaneous part of far-zone radiative energy flux in MH coordinates as a function of the separation between the two objects at some instant. Rather than
 writing the full expression for energy flux in MH coordinates we list here the difference 
to be added to the expression in SH coordinates [Eq.~\eqref{EFSH-FS}] 
to obtain the corresponding expression in MH coordinates. We have, 
\begin{equation}
\delta_{\rm (SH \rightarrow MH)}({d{\cal E}/dt})_{\rm inst}={16\over15}{c^5\over G}\,\nu^2\gamma^8
\left[-{110\over3}\,\nu\log\left({z\over z_0'}\right)+s\left({88\over3}\,\nu\log\left({z\over z_0'}\right)\right)+s^4\left({22\over3}\,\nu\log\left({z_i\over z_0'}\right)\right)
\right].
\label{EFMH-FS}
\end{equation}
\subsubsection{ Case II: Infall from infinity} 
By inserting  $s=z/z_i$ back in the Eq.~\eqref{EFMH-FS} and taking the limit when $z_i \rightarrow \infty$, the expression for ``$\delta_{\rm (SH \rightarrow MH)}{d{\cal E}/dt}$''
takes the form
\begin{equation}
\delta_{\rm (SH \rightarrow MH)}({d{\cal E}/dt})_{\rm inst}={16\over15}{c^5\over G}\,\nu^2\gamma^8
\left[-{110\over3}\,\nu\log\left({z\over z_0'}\right)
\right].
\label{EFMH-IS}
\end{equation}
\subsection{The 3PN instantaneous Energy Flux in ADM Coordinates }
\label{ADM}
In this section we  provide the expressions for instantaneous part of the energy flux in ADM coordinates which could be useful for the comparison with the numerical relativity results. One can  write the energy flux $\mathcal{F}$ in the ADM coordinates as
\begin{equation}
\mathcal{F}_{\rm ADM}=\mathcal{F}_{\rm SH}+\delta_{\rm (SH\rightarrow ADM)}\mathcal{F}\,,
\end{equation}
where $\mathcal{F}_{\rm SH}$ is given by Eqs.~\eqref{EF} and \eqref{3PNEF-inst} and the shift ``$\delta_{\rm (SH\rightarrow ADM)}\mathcal{F}$'' for the  head-on situation is
\begin{align} 
\label{EFSHtoADM}
\delta_{\rm (SH\rightarrow ADM)}\mathcal{F} =& -\frac{G^4\, m^5\, \nu
^2}{c^9\, z^5}\left\{-\frac{56}{15}\dot{z}^4\,\nu+\frac{G\, m}{z}\dot{z}^2 \left(\frac{4}{5}+\frac{88}{15}\,\nu\right)
\right\}
\nonumber\\& -\frac{G^4\, m^5\, \nu ^2}{c^{11}\, z^5}
\left\{ \dot{z}^6\left(\frac{628}{45}\,\nu-\frac{116}{45}\,\nu ^2\right)
 \right.\nonumber\\&\qquad +\frac{G\, m}{z}\dot {z}^4 \left(-\frac{512}{105}-\frac{24946}{315}\,\nu+\frac{386}{315}\,\nu^2\right)
\nonumber\\&\qquad 
+\frac{G^2\, m^2}{z^2} \dot {z}^2\left(\frac{36}{5}+\left[\frac{4568}{1575}-\frac{14}{5}\,\pi^2+\frac{1408}{45}\log \left(\frac{z}{z'_0}\right)\right]\nu+\frac{132}{5}\,\nu^2\right)
 \nonumber\\&\qquad \left.
+\frac{G^3\, m^3}{z^3}\left({16 \over 35}+\frac{128}{35}\, \nu-\frac{768}{35}\,\nu ^2 \right)\right\}\,.
\end{align} 
We have made use of Eq.~\eqref{GOtoHOC} to get Eq.~\eqref{EFSHtoADM} from the general expression for ``$\delta_{\rm (SH\rightarrow ADM)}\mathcal{F}$'', given by Eq.~(6.11) of~\cite{ABIQ07}.\\
\\
\subsubsection{Case I: Infall from a finite distance} 
Substituting for $\dot{z}$ in ADM coordinates in the expression for energy flux in ADM coordinates, we obtain the expression for the difference ``$\delta_{\rm (SH \rightarrow ADM)}{d{\cal E}/dt}$,'' which should be added to Eq.~\eqref{EFSH-FS} to obtain the instantaneous part of the energy flux in ADM coordinates. It reads 
\begin{eqnarray}
\delta_{\rm (SH \rightarrow ADM)}({d{\cal E}/dt})_{inst}&=&{16\over15}{c^5\over G}\,\nu^2\gamma^5\left\{
\left[-{5\over4}+{5\over2}\,\nu
+s\left(1-{39\over2}\,\nu\right)
+14\,s^2\,\nu
+s^3\left({1\over4}+3\,\nu\right)
\right]\gamma^2\right.\nonumber \\
& & \left.
+\sqrt{2}\sqrt{1-s}\left[{56\over15}\,\nu-{16\over5}\,s\,\nu\right]\gamma^{5/2}\right.\nonumber \\
& & \left.
+\left[{129\over14}+\left[{7355\over168}+{105\over32}\,\pi^2-{110\over3}\log\left({z\over z_0'}\right)\right]\nu+{482\over21}\,\nu^2\right.\right.\nonumber \\
& & \left.\left.
+s\left(-{145\over7}+\left[{20641\over210}-{21\over8}\,\pi^2+{88\over3}\log\left({z\over z_0'}\right)\right]\,\nu-{18875\over84}\,\nu^2\right)\right.\right.\nonumber \\
& & \left.\left.
+s^2\left({71\over7}-{11729\over42}\,\nu+{31027\over84}\,\nu^2\right)
+s^3\left(-{29\over7}+{7361\over84}\,\nu-{1873\over21}\,\nu^2\right)\right.\right.\nonumber \\
& & \left.\left.
+s^4\left({71\over14}+\left[{12917\over280}-{21\over32}\,\pi^2+{22\over3}\log\left({z_i\over z_0'}\right)\right]\,\nu-{405\over7}\,\nu^2\right)
\right]\gamma^3\right\}.
\label{EFADM-FS}
\end{eqnarray}
\subsubsection{ Case II: Infall from infinity}  
Inserting $s=z/z_i$ back in Eq.~\eqref{EFADM-FS} and taking the limit when $z_i \rightarrow \infty$ we obtain the expression for ``$\delta_{\rm (SH \rightarrow ADM)}({d{\cal E}/dt})$'' as
\begin{eqnarray}
\delta_{\rm (SH \rightarrow ADM)}({d{\cal E}/dt})_{\rm inst}&=&{16\over15}{c^5\over G}\,\nu^2\gamma^5\left\{
\left[-{5\over4}+{5\over2}\,\nu
\right]\gamma^2
+{56\,\sqrt{2}\over15}\,\nu\,\gamma^{5/2}\right.\nonumber \\
& & \left.
+\left[{129\over14}+\left[{7355\over168}+{105\over32}\,\pi^2-{110\over3}\log\left({z\over z_0'}\right)\right]\nu+{482\over21}\,\nu^2
\right]\gamma^3\right\}.
\label{EFADM-IS}
\end{eqnarray}
\section{Hereditary contributions in the flux For Head-on Collision}
\label{hereditary}
In this section we shall compute the hereditary contributions to the GW energy flux at 3PN order given by the Eqs.~\eqref{hered}, Eqs.~\eqref{Ftailhar}, and \eqref{Ftailtailhar}. The first hereditary contribution to the energy flux occurs at 1.5PN order and is due to GW tails caused by the interaction between mass quadrupole moment and the ADM mass of the source causing the spacetime curvature. This contribution is given by the first term in Eq.~\eqref{Ftailhar} where as the second term represents the subdominant tail at 2.5PN order caused due to interaction of a higher order multipole moment with the ADM mass of the source. Two cubic order tail terms, given by Eq.~\eqref{Ftailtailhar}, known as tails-of-tails and tail-squared occur at 3PN order and are caused due to the interaction of tails with ADM mass of the source and interaction of tails among themselves.

It should be evident from Eqs.~\eqref{Ftailhar} and \eqref{Ftailtailhar} that the computation of all terms would require only Newtonian order inputs except the mass-type quadrupolar tail term-- first term in Eq.~\eqref{Ftailhar} --which would include 1PN corrections. Note that the second term appearing in Eq.~\eqref{Ftailhar} and needed to be evaluated with Newtonian accuracy, does not contribute to the energy flux for the case of radial infall from infinity. The reason is that this term involves 4th and 6th derivatives of octupole moment [see Eq.~\eqref{MOHOC} for the expression] but one can check that, at the Newtonian order third and higher order derivatives of octupole moment vanish (after we substitute for the Newtonian $\dot{z}$ appearing in corresponding expressions), i.e. ${I_{ijk}}^{(n)}=0$ for $n > 2$, and hence the second integral would not contribute. In this case we only have the first term left in Eq.~\eqref{Ftailhar} which gives a hereditary contribution to GW energy flux at 2.5PN order. Computation of this term will require the knowledge of 1PN accurate expression for the quadrupole mass moment and 1PN accurate trajectory of the system. In addition to this the 1PN accurate expression for ADM mass would also be needed for the computations of tails at 2.5PN order. We have provided the 1PN trajectory in Sec.~\ref{inputs-hered} which will be used in computing the hereditary contributions. The second term of Eq. \eqref{Ftailhar} survives for the case of infall from a finite initial separation and thus must be taken into account while computing the tails at 2.5PN order for the finite initial separation case. As for the instantaneous part we will compute the hereditary contributions as well for two different situations, case I- infall from a finite distance and case II-infall from infinity.

It is important to note that, at 3PN order unlike the instantaneous part of the flux, the hereditary part is the same in all the three coordinate systems -- SH, MH and ADM since it involves the inputs which are at most required at 1PN order and are same in all three coordinate systems.

In addition to the inputs listed in Sec~\ref{inputs-hered} we also need 1PN accurate expressions for mass quadrupole moment and for ADM mass. The mass quadrupole moment in terms of the variables $z$ and $\dot z$ is given by Eq.~\eqref{MQHOC}. In the 1PN limit it reads, 
\begin{equation}
I_{ij} = \nu\, m\,z^2\left\{1+\frac{1}{c^2}\left[\dot {z}^2\left(\frac{9}{14} - \frac{27}{14}\,\nu\right)+\frac{G\,m}{z}\left(-\frac{5}{7} + \frac{8}{7}\,\nu\right)
\right]
\right\}n_{\langle i}n_{j \rangle}\,,\label{MQ1PN}
\end{equation}
The relation between the ADM mass $M$ and total mass $m=m_1+m_2$ is given by,
\begin{equation}
M=m\left[1+\frac{\nu}{c^2}\left(\frac{\dot{z}^2}{2}-\frac{G\,m}{z}\right)\right]\,,
\label{ADM-m}
\end{equation}
where $\dot{z}$ is given by Eq.~\eqref{zdot-square-SH-FS} and is needed to be just Newtonian accurate.\\
\\
\subsection{Case I: Infall from a finite distance}
\label{subsec:FheredFS}
In this case the 1PN accurate expression for quadrupole moment and Newtonian accurate expression for octupole moment take the form,
\begin{subequations}
\begin{equation}
I_{ij}=m\,z^2\,\nu\left\{1+{G\,m \over c^2 \,z}\left[{4\over 7}-{19\over7}\,\nu +s\left(-{9\over 7}+{27\over 7}\,\nu\right)\right]\right\}\,n_{\langle i}n_{j \rangle}\,,\label{MQ1PNz}
\end{equation}
\begin{equation}
I_{ijk}=-\nu\,m\,z^3\,\sqrt{1-4\,\nu}\,n_{\langle i}n_{j}n_{k\rangle}\,.\label{MONewtz}
\end{equation}
\end{subequations}
The relation between ADM mass and total mass at 1PN order is given as,
\begin{equation}
M=m\left[1-{G\, m\,\nu \over c^2\,z_{i}}\right]\,.
\label{M-m-FS}
\end{equation}       
From the above expression the ADM mass $M$ is independent of $z$ which is
consistent with the constancy of $M$ and the recognition of the expression
of the initial Newtonian energy $-G m \nu/z_i$.\\
Hereditary terms will involve two integrals, which are given as
\begin{subequations}
\begin{equation}
I_{\rm tail}^{\rm Quadrupole}=\int_{u(z_{i})}^{u}d\tau\,I^{(5)}_{ij}(\tau)\biggl[\log\left(\frac{c}{2\,z}\,(u-\tau)\right)
+\frac{11}{12}\biggr]\,,\label{tailintfs}
\end{equation}
\begin{equation}
I_{\rm tail}^{\rm Octupole}=\int_{u(z_{i})}^{u}d\tau\,I^{(6)}_{ijk}(\tau)\biggl[\log\left(\frac{c}{2\,z}\,(u-\tau)\right)
+\frac{97}{60}\biggr]\,,\label{tailint2fs}
\end{equation}
\begin{equation}
I_{\rm tail(tail)}=\int_{u(z_i)}^{u}d\tau\,I^{(6)}_{
ij}(\tau)\biggl[\log^2\left(\frac{
c}{2\,z}\,(u-\tau)\right)+\frac{57}{70}\log\left(\frac{
c}{2\,z}\,(u-\tau)\right)+\frac{124627}{44100}\biggr]\,.\label{taioftailintfs}
\end{equation}
\end{subequations}
After evaluating the integrals we get
\begin{subequations}
\label{int_value_FS}
\begin{eqnarray}
I_{\rm tail}^{\rm Quadrupole}&=&{G^2\,m^3\,\nu \over z^4}\left\{\biggl[{55 \over 6}
-5\, \log \left(8\,\gamma \right)
+s\left(-{22 \over 3}
+4\,\log \left(8\,\gamma\right)\right)
+s^4\left(-{11 \over 6}
+\log (8\,\gamma)+2\,{\rm Int1}(s)\right)
\biggr]\right.\nonumber \\ & & \left.
+\biggl[{17 \over 3}\left(1-\nu\right)\Big(-11+6\,\log\left(8\,\gamma\right)\Big)
+s\left[\left(-{19\over 2}+{67\over 6}\,\nu \right)\Big(-11+6\,\log\left(8\,\gamma\right)\Big)\right]\right.\nonumber \\ & & \left.
+s^2\left[\left({80\over 21}-{107 \over 21}\,\nu\right)\Big(-11+6\,\log\left(8\,\gamma\right)\Big)\right]
+s^5\left[
\left({1\over 42}-{17\over 42}\,\nu \right)\Big(-11+6\,\log\left(8\,\gamma\right)\Big)\right.\right.\nonumber \\ & & \left.\left.
-2\,{\rm Int20}(s)+{\rm Int30}(s)-{\rm Int40}(s)
+\nu \left(-2\,{\rm Int21}(s)+{\rm Int31}(s)-{\rm Int41}(s)\right)
\right]
\biggr]\,\gamma \right\}\,n_{\langle i}n_{j \rangle},
\label{int_tail_FS}
\end{eqnarray}
\begin{eqnarray}
I_{\rm tail}^{\rm Octupole}&=&\frac{G^{5/2}\,m^{7/2}\,\nu}{z^{9/2}}\,\sqrt{1-4\,\nu}\left\{\sqrt{2}\sqrt{1-s}\left[s \left(-\frac{194}{5}+12 \log(8 \gamma)\right)-\frac{12 s^{9/2}}{\sqrt{1-s}}{\rm Int7}(s)\right]\right\}n_{\langle i}n_{j}n_{k \rangle},
\label{int2_tail_FS}
\end{eqnarray}
\begin{eqnarray}
I_{\rm tail(tail)}&=&{G^{5/2}\,m^{7/2}\,\nu \over z^{11/2}}\,\sqrt{2}\,\sqrt{1-s}\left\{{249254 \over 2205}-{114 \over 7}\,\log \left(8\,\gamma \right)+10\,\log^2\left(8\,\gamma\right)
+s\,\left(-{249254 \over 3675}+{342 \over 35}\,\log \left(8\,\gamma\right)
\right.\right.\nonumber \\ & & \left.\left.
-6\,\log^2 \left(8\,\gamma\right)\right)
+{s^{11/ 2} \over \sqrt {1-s}}\,\left(2\, {\rm Int5}(s)+{57 \over 35}\,{\rm Int6}(s)-2\,{\rm Int6}(s)\, \log \left(8\,\gamma\right)
\right)\right\}\,n_{\langle i}n_{j \rangle}\,.
\label{int_tailoftail_FS}
\end{eqnarray}
\end{subequations}
where Int1$(s)$, Int20$(s)$, Int21$(s)$, Int30$(s)$,  Int31$(s)$, 
Int40$(s)$, and Int41$(s)$ appearing in Eq.~\eqref{int_tail_FS} and Int7$(s)$ appearing in Eq.~\eqref{int2_tail_FS} are given by
\begin{subequations}
\begin{equation}
{\rm Int1}(s) = 4\int^1_s
\biggl(\frac{5-3y}{y^5}\biggr)\,\log\biggl[s^{-3/2}\Big(g(s)-g(y)\Big)\biggr]\,dy,
\label{Int1s}
\end{equation}
\begin{equation}
{\rm Int20}(s) = \int^1_s
\biggl(\frac{1540 -1876\,y+522\,y^2}{7\,y^6}
\biggr)\log\biggl[s^{-3/2}\,\Big(g(s)-g(y)\Big)\biggr]\,dy,
\label{Int20s}
\end{equation}
\begin{equation}
{\rm Int21}(s) = \int^1_s
\biggl(\frac{-1365+2296\,y-831\,y^2}{7\,y^6}
\biggr)\log\biggl[s^{-3/2}\,\Big(g(s)-g(y)\Big)\biggr]\,dy,
\label{Int21s}
\end{equation}
\begin{equation}
{\rm Int30}(s) = 4\int^1_s
\biggl(\frac{\left(5-3\,y\right)\,\left(5-\,y\right)}{y^6}
\biggr)\log\biggl[s^{-3/2}\Big(g(s)-g(y)\Big)\biggr]\,dy,
\label{Int30s}
\end{equation}
\begin{equation}
{\rm Int31}(s) = 2\int^1_s
\biggl(\frac{\left(5-3\,y\right)\,\left(-5+9y\right)}{y^6}
\biggr)\log\biggl[s^{-3/2}\Big(g(s)-g(y)\Big)\biggr]\,dy,
\label{Int31s}
\end{equation}
\begin{equation}
{\rm Int40}(s) = 4\int^1_s
\biggl(\frac{5-3y}{y^5}\biggr)\left(\frac{h_0(s)-h_0(y)}{g(s)-g(y)}\right)\,dy,
\label{Int40s}
\end{equation}
\begin{equation}
{\rm Int41}(s) = 4\int^1_s
\biggl(\frac{5-3y}{y^5}\biggr)\left(\frac{h_1(s)-h_1(y)}{g(s)-g(y)}\right)\,dy.
\label{Int41s}
\end{equation}
\begin{equation}
{\rm Int7}(s) = \int^1_s
\biggl(\frac{7-6y}{y^{9/2}\sqrt{1-y}}\biggr)\,\log\biggl[s^{-3/2}\Big(g(s)-g(y)\Big)\biggr]\,dy,
\label{Int7s}
\end{equation}
\end{subequations}
and Int5$(s)$ and Int6$(s)$ appearing in Eq.~\eqref{int_tailoftail_FS} are  given by
\begin{subequations}
\begin{equation}
{\rm Int5}(s) = \int^1_s
{1 \over \sqrt {1-y}}\left(\frac{110-154\,y+48\,y^2}{y^{13/2}}\right)\log ^2\biggl[s^{-3/2}\Big(g(s)-g(y)\Big)\biggr]\,dy\,,
\label{Int5s}
\end{equation}
\begin{equation}
{\rm Int6}(s) = \int^1_s
{1 \over \sqrt {1-y}}\left(\frac{110-154\,y+48\,y^2}{y^{13/2}}
\right)\,\log \biggl[s^{-3/2}\Big(g(s)-g(y)\Big)\biggr]\,dy.
\label{Int6s}
\end{equation}
\end{subequations}
With Eq.~\eqref{int_value_FS} we can write the various pieces of the hereditary contributions to GW energy flux given by Eq.~\eqref{hered} as 
\begin{eqnarray}
{\cal{F}}_\mathrm{tail}
&=& { 16 \over 15}\,\frac{c^5}{G}\,\nu^2\,\gamma^5
\left \{\sqrt{2}\sqrt {1-s} \biggl[{55 \over 6}-5\,\log \left(8\,\gamma\right)+s\left(-{22 \over 3}+4\,\log \left(8\,\gamma\right)\right)
+s^4\left(-{11 \over 6}+2\, {\rm Int1}(s)
+\log (8\,\gamma) \right)
\biggr]\gamma^{3/2}\right.\nonumber \\ & & \left.
+\sqrt {2}\sqrt {1-s} \biggl[-{7601 \over 84}+{16577\over 168}\,\nu 
+\left(\frac{691}{14}-\frac{1507}{28}\,\nu \right)
\,\log \left(8\,\gamma \right)
+s\left(\frac{4895}{28}-\frac{35365}{168}\,\nu
\right.\right.\nonumber \\ & & \left.\left.
+\left(-{1335 \over 14}+{3215\over 28}\,\nu\right)
\,\log \left(8\,\gamma\right)\right)
+s^2\left(-{8027 \over 105}+{18511 \over 210}\,\nu+\left({298 \over 7}-{361 \over 7}\,\nu\right)
\, \log \left(8 \,\gamma\right)\right)
+s^4\left({473 \over 84}-{407\over 56}\,\nu \right.\right.\nonumber \\ & & \left.\left.
+\left(-{43 \over 7}
+{111\over 14}\,\nu\right)\,{\rm Int1}(s)
+\left(-{43 \over 14}+{111 \over 28}\,\nu\right)\,\log \left(8\,\gamma\right)\right)
+s^5\left(-{275 \over 28}+{2717\over 168}\,\nu 
+\left(\frac{73}{7}-\frac{179}{14}\,\nu 
\right)\,{\rm Int1}(s)\right.\right.\nonumber \\ & & \left.\left.
-2\,{\rm Int20}(s) +{\rm Int30}(s)-{\rm Int40}(s)
+\nu\left(-2\,{\rm Int21}(s) +{\rm Int31}(s)-{\rm Int41}(s)\right)
+\left(\frac{75}{14}-\frac{247}{28}\,\nu\right)\,\log \left(8\,\gamma \right)\right)
\right.\nonumber\\ && \left.+{s^{11/2}\over \sqrt{1-s}}\left({8\over 7}-{32\over 7}\,\nu\right){\rm Int7(s)}
\biggr]\gamma^{5/2}
 \right \}\,,\nonumber\\
&&
\label{Tail_FS}
\end{eqnarray}
\begin{eqnarray}
{\cal{F}}_\mathrm{tail(tail)}
&=& { 16 \over 15}\,\frac{c^5}{G}\,\nu^2\,\gamma^5
\left\{\biggl [\left(\frac{498508}{2205}
-{228 \over 7}\log \left(8\,\gamma \right)
+20 \log ^2\left(8\,\gamma \right)
-{1712 \over 21}\log \left({z \over z_0}\right)\right)
+s \left(-\frac{3988064}{11025}
\right.\right.\nonumber \\ & & \left.\left.
+{1824 \over 35}\log \left(8\,\gamma\right)
-32 \log ^2\left(8\,\gamma \right)
+{13696 \over 105}\log \left({z \over z_0}\right)\right)
+s^2 \left(\frac{498508}{3675}
-{684 \over 35}\log \left(8\,\gamma\right)
+12\,\log ^2\left(8\,\gamma\right)
\right.\right.\nonumber \\ & & \left.\left.
-{1712 \over 35}\log \left({z \over z_0}\right)\right)
+s^{11/2}\sqrt {1-s} \left(4\,{\rm Int5}(s)+\left({114 \over 35}
-4\,\log \left(8\,\gamma \right)\right) {\rm Int6}(s)
\right)\biggr]\gamma^3\right\}\,,
\label{Tailoftail_FS}
\end{eqnarray}
\begin{eqnarray}
{\cal{F}}_\mathrm{(tail)^2}
&=& { 16 \over 15}\,\frac{c^5}{G}\,\nu^2\,\gamma^5
\left\{\biggl [\frac{3025}{72}-\frac{275}{6}\, \log \left(8\,\gamma \right)
+\frac{25 }{2}\,\log ^2\left(8\,\gamma \right)
+s\left(-\frac{605}{9}
+\frac{220}{3}\, \log \left(8\,\gamma \right)
-20\,\log ^2\left(8\,\gamma \right)\right)
\right.\nonumber \\ & & \left.
+s^2 \left(\frac{242}{9}
-\frac{88}{3} \log \left(8\,\gamma \right)
+8\, \log ^2\left(8\,\gamma \right)\right)
+s^{4} \left(-\frac{605}{36}+{55 \over 3}\, {\rm Int1}(s)
+\left(\frac{55}{3}
-10\,{\rm Int1}(s)\right) \log \left(8\,\gamma \right)
\right.\right.\nonumber \\ & & \left.\left.
-5 \log ^2\left(8\,\gamma \right)
\right)
+s^{5} \left(\frac{121}{9}-\frac{44}{3}{\rm Int1}(s)
+\left(-\frac{44}{3}
+8\, {\rm Int1}(s)\right) \log \left(8\,\gamma \right)
+4\, \log ^2\left(8\,\gamma \right)\right)
\right.\nonumber \\ & & \left.+
s^{8} \left({121 \over 72}-{11 \over 3}{\rm Int1}(s)+2\, [{\rm Int1}(s)]^2
+\left(-\frac{11}{6} 
+2\,{\rm Int1}(s)\right) \log \left(8\,\gamma \right)
+\frac{1}{2}\log ^2\left(8\,\gamma \right)
\right)
\biggr]\gamma^3\right\}\,.
\label{Tailsquared_FS}
\end{eqnarray}
Combining Eqs.~\eqref{hered}, \eqref{Tail_FS}, \eqref{Tailoftail_FS} and \eqref{Tailsquared_FS} now we can write the hereditary contribution at 3PN order involving contribution from tails, tail-of-tail and tail squared terms as 
\begin{eqnarray}
\label{Fhered-FS} {\cal F}_\mathrm{hered} &=&  { 16 \over 15} \frac{c^5}{G}\,\nu^2\, \gamma^5\left\{\sqrt{2}\sqrt {1-s} \biggl[{55 \over 6}-5\,\log \left(8\,\gamma\right)+s\left(-{22 \over 3}+4\,\log \left(8\,\gamma\right)\right)
+s^4\left(-{11 \over 6}+2\, {\rm Int1}(s)
+\log (8\,\gamma) \right)
\biggr]\gamma^{3/2}
\right.\nonumber \\ 
& & \left.
+\sqrt {2}\sqrt {1-s} \biggl[-{7601 \over 84}+{16577\over 168}\,\nu 
+\left(\frac{691}{14}-\frac{1507}{28}\,\nu \right)
\,\log \left(8\,\gamma \right)
+s\left(\frac{4895}{28}-\frac{35365}{168}\,\nu
\right.\right.\nonumber \\ 
& & \left.\left.
+\left(-{1335 \over 14}+{3215\over 28}\,\nu \right)
\,\log \left(8\,\gamma\right)\right)
+s^2\left(-{8027 \over 105}+{18511 \over 210}\,\nu+\left({298 \over 7}-{361 \over 7}\,\nu\right)
\, \log \left(8 \,\gamma\right)\right)
+s^4\left({473 \over 84}-{407\over 56}\,\nu 
\right.\right.\nonumber \\ 
& & \left.\left.
+\left(-{43 \over 7}+{111\over 14}\,\nu\right)\,{\rm Int1}(s)
+\left(-{43 \over 14}+{111 \over 28}\,\nu\right)
\, \log \left(8\,\gamma\right)\right)
+s^5\left(-{275 \over 28}+{2717 \over 168}\,\nu
+\left(\frac{73}{7}-\frac{179}{14}\,\nu 
\right)\,{\rm Int1}(s)\right.\right.\nonumber \\ 
& & \left.
-2\,{\rm Int20}(s)+{\rm Int30}(s)-{\rm Int40}(s)
+\nu\left(-2\,{\rm Int21}(s)+{\rm Int31}(s)-{\rm Int41}(s)\right)
+\left(\frac{75}{14}-\frac{247}{28}\,\nu\right)\,\log \left(8\,\gamma \right)\right)
\nonumber \\ & &
+{s^{11/2}\over \sqrt{1-s}}\left({8\over 7}-{32\over 7}\,\nu\right){\rm Int7(s)}
\biggr]\gamma^{5/2}
\nonumber \\ 
& & 
+\biggl[\frac{4729189}{17640}
-{3293 \over 42}\log \left(8\,\gamma\right)
+{65 \over 2}\log^2\left(8\,\gamma\right)
-{1712 \over 21}\log \left({z \over z_0}\right)
+s \left(-\frac{4729189}{11025}
+\frac{13172}{105}\log \left(8\,\gamma \right)
\right.\nonumber\\ 
& &\left.
-52 \,\log ^2\left(8\,\gamma \right)+\frac{13696 }{105}\log \left({z \over z_0}\right)
\right)
+s^{2}\left(\frac{1791974}{11025}
-\frac{5132}{105}\, \log \left(8\,\gamma \right)
+20\, \log ^2\left(8\,\gamma\right)
-\frac{1712}{35}\log \left({z \over z_0}\right)\right)
\nonumber\\ 
& &
+s^{4} \left(-\frac{605}{36}+{55 \over 3}\, {\rm Int1}(s)
+\left(\frac{55}{3}
-10\,{\rm Int1}(s)\right) \log (8\,\gamma )
-5 \log ^2(8\,\gamma )\right)
+s^{5} \left(\frac{121}{9}-\frac{44}{3}{\rm Int1}(s)
\right.\nonumber \\ 
&&\left.
+\left(-\frac{44}{3}
+8\, {\rm Int1}(s)\right) \log \left(8\,\gamma \right)
+4\, \log ^2\left(8\,\gamma \right)\right)
+s^{11/2}\sqrt {1-s} \left(4\,{\rm Int5}(s)
+\left({114 \over 35}
-4\,\log \left(8\,\gamma \right)\right) {\rm Int6}(s)\right)
\nonumber\\
&&\left.
+s^{8} \left({121 \over 72}-{11 \over 3}{\rm Int1}(s)+2\, [{\rm Int1}(s)]^2
+\left(-\frac{11}{6} 
+2\,{\rm Int1}(s)\right) \log \left(8\,\gamma \right)
+\frac{1}{2}\,\log ^2\left(8\,\gamma \right)
\right)
\biggr]{\gamma}^{3}
\right\}.
\end{eqnarray}
As we can see the above equation still has some dependence on the arbitrary scale $z_0$ at 3PN order. Recall, the presence of a logarithmic dependence on $z_0$ in the instantaneous contribution to the energy flux at 3PN order. The term appearing in the hereditary contribution exactly cancels with similar terms present in instantaneous flux expression for energy flux and thus the total flux becomes independent of the arbitrary length scale $z_0$ as expected.\\
\subsection{Case II: Infall from infinity}
\label{subsec:FheredIS}
The 1PN accurate expression for quadrupole moment reads
\begin{equation}
I_{ij}=m\,z^2\,\nu\left[1+{G\,m \over c^2 \,z}\left({4\over 7}-{19\over 7}\,\nu \right)\right]\,n_{\langle i}n_{j \rangle}\,.
\end{equation}
It is evident from Eq.~\eqref{zdot-IS}, the relation connecting ADM mass $M$ and total mass $m$ [Eq.~\eqref{ADM-m}] reduces to
\begin{equation}
 M=m\,.
\end{equation}
This is consistent with the earlier comment and corresponds to  energy
vanishing initially.
In order to compute the hereditary contribution first we need to evaluate the two integrals appearing in Eqs.~\eqref{Ftailhar} and \eqref{Ftailtailhar}. The integral associated with the first term of Eq.~\eqref{Ftailhar} is
\begin{equation}
I_{\rm tail}=\int_0^{+\infty}d\tau\,I^{(5)}_{ij}(u-\tau)\biggl[\log\left(\frac{c\,\tau}{2\,z}\right)
+\frac{11}{12}\biggr]\,,\label{tailintis}
\end{equation}
and the integral appearing in the first term of Eq.~\eqref{Ftailtailhar} is
\begin{equation}
I_{\rm tail(tail)}=\int_{0}^{+\infty}d\tau\,I^{(6)}_{
ij}(u-\tau)\biggl[\log^2\left(\frac{
c\,\tau}{2\,z}\right)+\frac{57}{70}\log\left(\frac{
c\,\tau}{2\,z}\right)+\frac{124627}{44100}\biggr]\,.\label{taioftailintis}
\end{equation}
Having all the relevant inputs at the required PN order the value of these integrals read,
\begin{subequations}
\label{int_value}
\begin{eqnarray}
I_{\rm tail}&=&\frac{G^2\, m^3\, \nu }{z^4}\left\{ \left[-{71 \over 6}-{5\over \sqrt {3}}\,\pi-5 \log \left({2 \over 3}\,\gamma \right)\right] \right.\nonumber \\ & & \left.
+\left[-{2497 \over 21}+{166 \over \sqrt {3}}\,\pi+\left[-{2161\over 42} -22\,\sqrt {3}\,\pi\right]\nu+34\,(1-\nu)\,\log\left({2 \over 3}\, \gamma\right)
\right]\,\gamma
\right\}\, n_{\langle i}n_{j \rangle}, 
\end{eqnarray}
\begin{eqnarray}
I_{\rm tail(tail)}&=&\,{G^{5/2}\,m^{7/2}\,\nu \over \sqrt{2}\,z^{11/2}}
\left\{\frac{4894237}{8820}+\frac{386\, \sqrt{3}}{7}\,\pi+20\, \pi ^2+\left({1158 \over 7}+{40\over \sqrt{3}}\,\pi\right)\,\log \left({2 \over 3}\, \gamma \right)\right.\nonumber\\ & &\left.+20\,\log^2 \left({2 \over 3}\, \gamma\right)+80\, \psi^{(1)}\left(\frac{11}{3}\right) 
\right\}\, n_{\langle i}n_{j \rangle}\,,
\label{tailoftailint}
\end{eqnarray}
\end{subequations}
where the quantity  $\psi ^{(1)}\left(\frac{11}{3}\right)$ appearing in Eq.~\eqref{tailoftailint} is a PolyGamma function whose numerical value is 0.313 25.
(Of course, formally they correspond to $s\rightarrow 0$ case of
the previous section.) \\
\\
Using Eq.~\eqref{int_value} in Eqs.~\eqref{Ftailhar}-\eqref{Ftailtailhar} we write various pieces of hereditary contribution given by Eq.~\eqref{hered} as 
\begin{eqnarray}
{\cal{F}}_\mathrm{tail}
&=& { 16 \over 15} \frac{c^5}{G} \nu^2 \gamma^5
\left \{\sqrt{2} \biggl[-{71 \over 6}-{5 \over \sqrt {3}}\,\pi
-5 \log \left({2 \over 3}\gamma \right)\biggr]{\gamma}^{3/2} \right.\nonumber \\ & & \left.
+\sqrt {2} \biggl[-\frac{6935}{84}+\frac{2539}{14\, \sqrt{3}}\,\pi+\left(-\frac{16525}{168}-\frac{801\,\sqrt{3}}{28}\,\pi\right)\nu
+\left({691 \over 14}
-\frac{1507}{28}\,\nu\right)\log \left({2 \over 3}\gamma \right)
\biggr]\gamma ^{5/2}
\right \}\,.
\label{Tail_IS}
\end{eqnarray}
As mentioned earlier the 2PN accurate energy flux has been given in~\cite{SPW95} which involves the hereditary contribution to the energy flux at 1.5PN order. 
On comparing our results
[1.5PN term in Eq.~\eqref{Tail_IS} above with coefficient -5] 
and [Eq.~(2.31) of~\cite{SPW95} with coefficient -15] for the contribution due to dominant tail we find a mismatch.  
This apparent discrepancy is a gauge-artifact arising from the difference
in our choice of $u=t-r/c$ in contrast to the choice in SPW~\cite{SPW95}
$u_{\rm SPW}=t-r/c-(2\,G\,M/ c^3)\log\left(c^2\,r / G\,m\right)$.
We have checked that once we adopt 
the SPW definition of retarded time in harmonic coordinates $u_{\rm SPW}$,
our result also leads to the  coefficient $-15$  as  in ~\cite{SPW95}.        
This difference serves to remind us that the representation of the energy flux in terms of $\gamma$ is {\it not} gaugeinvariant.
\begin{eqnarray}
 \,\,\,\,{\cal{F}}_\mathrm{tail(tail)}
&=& { 16 \over 15}\,\frac{c^5}{G}\, \nu^2\, \gamma^5
\left \{\biggl[\frac{4894237}{8820}+\frac{386\, \sqrt{3}}{7}\,\pi+20\,\pi^2
-\frac{1712}{21}\,\log\left(\frac{z}{z_0}\right)+\left({1158 \over 7}+\frac{40}{\sqrt{3}}\,\pi\right)\,\log \left({2 \over 3}\gamma \right)\right.\nonumber \\ & & \left.+20\, \log ^2\,\left({2 \over 3}\,\gamma \right)
+80\, \psi^{(1)}\left(\frac{11}{3}\right)\biggr]{\gamma}^{3}  \right \}\,,
\label{Tailoftail_IS}
\end{eqnarray}
\begin{eqnarray}
{\cal{F}}_\mathrm {(tail)^2}
&=& { 16 \over 15}\frac{c^5}{G}\,\nu^2\,\gamma^5
\left \{\biggl[\frac{5041}{72}+\frac{355}{6\, \sqrt{3}}\,\pi+\frac{25}{6}\,\pi^2
+\left(\frac{355}{6}
+\frac{25}{\sqrt{3}}\,\pi\right) \log \left({2 \over 3}\,\gamma \right)+\frac{25}{2}\,\log ^2\left({2 \over 3}\,\gamma \right)
\biggr]{\gamma}^{3}  \right \}\,.
\label{Tailsquared_IS}
 \end{eqnarray} 
Now we can write the total hereditary contribution up to 3PN order to the energy flux as 
\begin{eqnarray}
\label{Fhered-IS} {\cal F}_\mathrm{hered} &=&  { 16 \over 15} \frac{c^5}{G}\,\nu^2\,\gamma^5\left\{\sqrt{2}\left[-
\frac{71}{6}-\frac{5}{\sqrt{3}}\,\pi
-5\,\log\left({2 \over 3}\,\gamma \right)
\right]{\gamma}^{3/2}\right.\nonumber\\ & &\left.
+\sqrt{2}\biggl[-\frac{6935}{84}+\frac{2539}{14\, \sqrt{3}}\,\pi+\left(-\frac{16525}{168}-\frac{801\,\sqrt{3}}{28}\,\pi\right)\nu
+\left({691\over 14}-\frac{1507}{28}\,\nu\right)\,\log \left({2 \over 3}\,\gamma \right)
\biggr]{\gamma}^{5/2}
\right.\nonumber\\ & &\left.
+\biggl[\frac{11023519}{17640}
+\frac{9433}{42 \,\sqrt{3}}\,\pi+\frac{145}{6}\,\pi^2-\frac{1712}{21} \log \left(\frac{z}{z_0}\right)
+\left(\frac{9433}{42}+\frac{65}{\sqrt{3}}\,\pi\right)\,\log \left({2 \over 3}\,\gamma \right)
\right.\nonumber\\ & &\left.
+\frac{65}{2} \log ^2\left({2 \over 3}\,\gamma\right)
+80 \psi^{(1)}\left(\frac{11}{3}\right)
\biggr]{\gamma}^{3}
\right\}\,.
\end{eqnarray}
The presence of the arbitrary scale $z_0$ in the above expression is similar to the one already noted in Eq.~\eqref{Fhered-FS} and will disappear from the final expression for energy flux.\\ 
\section{The complete 3PN Energy Flux for Head-on Situation}\label{completeflux}
\subsection{Case I: Infall from finite a distance}
\label{completefluxFS}
Having computed both the instantaneous and the hereditary contributions to the energy flux at 3PN order for head-on situation we are now ready to write the complete 3PN far-zone energy flux due to head-on infall of two compact objects with arbitrary mass ratios. Since the ADM coordinates are
independent of gauge-dependent logarithms they  may be
better suited for comparison with numerical relativity results,
and we exhibit the complete 3PN accurate energy flux expression 
in these coordinates obtained by adding the hereditary part [Eq.~\eqref{Fhered-FS}] and instantaneous part [Eq.~\eqref{EFADM-FS}] of the energy flux. The final result is
\begin{eqnarray}
\left( {d{\cal E} \over dt}\right)_{\rm ADM} 
&=& {16 \over 15}\frac{c^5}{G}\nu^2 \gamma^5
\left \{ 1 - s +\left[-{43 \over 7} +{111 \over 14}\,\nu
+s \left({116 \over 7} -{131\over 7}\,\nu\right ) + s^2\left(- {71 \over 7} +{135\over 14}\,\nu \right)
\right ]{\gamma}
\right.\nonumber \\
& & \left.
+\sqrt{2}\sqrt {1-s} \biggl[{55 \over 6}
-5\, \log \left(8\, \gamma \right)
+s\left(-{22 \over 3}
+4\,\log \left(8\,\gamma\right) \right)
+s^4\left(-{11 \over 6}+2\, {\rm Int1}(s)
+\log \left(8\,\gamma\right) \right)
\biggr]\gamma^{3/2}\right.\nonumber \\ & & \left.
+ \left [-\frac{4643}{108}-{713 \over 36}\,\nu+{112 \over 3}\,\nu^2
+s\left(-\frac{4282}{189}+{28505 \over 126}\,\nu-{2864 \over 21}\,\nu^2\right) 
+s^2\left(\frac{1870}{21}-{5251\over 18}\,\nu +{8800 \over 63}\, \nu^2\right) 
\right.\right.\nonumber \\
& & \left.\left.
+ s^3\left(-\frac{329}{12}+{1219 \over 12}\,\nu-{872\over 21}\,\nu^2\right) 
\right]
{\gamma^2}
+\sqrt {2} \sqrt {1-s}\biggl[-{7601 \over 84}+{11651\over 120}\,\nu 
+\left(\frac{691}{14}-\frac{1507}{28}\,\nu\right)
\,\log \left(8\,\gamma \right)
\right.\nonumber \\ & & \left.
+s\left(\frac{4895}{28}-\frac{173113}{840}\,\nu
+\left(-{1335 \over 14}+{3215 \over 28}\,\nu\right)
\,\log \left(8\,\gamma\right)\right)
\right. \nonumber \\ & & \left.
+s^2\left(-{8027 \over 105}+{18031 \over 210}\,\nu
+\left({298 \over 7}-{361\over 7}\,\nu \right)
\,\log\left(8 \,\gamma\right)\right)\right.\nonumber \\ & & \left.
+s^4\left({473 \over 84}-{407 \over 56}\,\nu+\left(-{43 \over 7}+{111\over 14}\,\nu\right)\,{\rm Int1}(s)+\left(-{43 \over 14}+{111 \over 28}\,\nu\right)
\, \log \left(8\,\gamma\right)\right)
\right.\nonumber \\
& & 
+s^5\left(-{275 \over 28}+{2717\over 168}\,\nu
+\left(\frac{73}{7}-\frac{179}{14}\,\nu 
\right)\,{\rm Int1}(s)
-2\,{\rm Int20}(s) +{\rm Int30}(s)-{\rm Int40}(s)
\right.\nonumber\\
&&\left.
+\nu\left(-2\,{\rm Int21}(s) +{\rm Int31}(s)-{\rm Int41}(s)\right)
+\left(\frac{75}{14}-\frac{247}{28}\,\nu\right)\,\log \left(8\,\gamma\right)\right)
+{s^{11/2}\over \sqrt{1-s}}\left({8\over 7}-{32\over 7}\,\nu\right){\rm Int7(s)}
\biggr]{\gamma^{5/2}}
\nonumber \\
& & \left.
+\left[\frac{155961373}{582120}+\left[-\frac{5467459}{5544}-\frac{1125}{32}\,\pi^2\right]\nu+\frac{1865377}{5544}\,\nu^2+\frac{231520}{2079}\,\nu^3-{3293 \over 42}\log \left(8\,\gamma\right)
+{65 \over 2}\log^2\left(8\,\gamma\right)
\right.\right.\nonumber \\
& & \left.\left.
+ s\,\biggl(-\frac{8236351}{40425}+\left[
\frac{5608279}{8316}+\frac{819}{16}\,\pi^2\right]\nu+
\frac{1302751}{1386}\,\nu^2-\frac{1366537}{2079}\,\nu^3
+\frac{13172}{105}\log \left(8\,\gamma \right)
-52 \,\log ^2\left(8\,\gamma \right)
\biggr)\right.\right.\nonumber \\
& & \left.\left.
+s^2\biggl(-\frac{74317681}{363825}+\left[\frac{1312231}{924}-\frac{123}{8}\,\pi^2\right]\nu-\frac{252311}{84}\,\nu^2+\frac{304961}{297}\,\nu^3
-\frac{5132}{105}\, \log \left(8\,\gamma \right)
+20\, \log ^2\left(8\,\gamma\right)
\biggr)\right.\right.\nonumber \\
& & \left.\left.
+s^3\biggl(\frac{207379}{1155}-\frac{4141933}{2772}\,\nu+\frac{6033829}{2772}\,\nu^2-\frac{406498}{693}\,\nu^3\biggr)
+s^4\biggl(-\frac{77347}{2772}+\left[\frac{74357}{264}-\frac{21}{32}\,\pi^2\right]\nu-\frac{848843}{1848}\,\nu^2
\right.\right.\nonumber \\ & & \left.\left.
+\frac{8076}{77}\,\nu^3+{55 \over 3}\,{\rm Int1}(s)
+\left(\frac{55}{3}
-10\,{\rm Int1}(s)\right) \log (8\,\gamma )
-5 \log ^2(8\,\gamma )\biggr)
+s^{5} \biggl(\frac{121}{9}-\frac{44}{3}{\rm Int1}(s)
\right.\right.\nonumber \\
& & \left.\left.
+\left(-\frac{44}{3}
+8\, {\rm Int1}(s)\right) \log \left(8\,\gamma \right)
+4\, \log ^2\left(8\,\gamma \right)
\biggr)
+s^{11/2}\sqrt {1-s} \left(4\,{\rm Int5}(s)+\left({114 \over 35}
-4\,\log \left(8\,\gamma \right)\right) {\rm Int6}(s)
\right)\right.\right.\nonumber \\ & & \left.\left.
+s^{8}\biggl({121 \over 72}-{11 \over 3}{\rm Int1}(s)+2\, [{\rm Int1}(s)]^2
+\left(-\frac{11}{6}
+2\,{\rm Int1}(s)\right) \log \left(8\,\gamma \right)
+\frac{1}{2}\,\log ^2\left(8\,\gamma \right)
\biggr)
\right]{\gamma^3} \right \}.
\label{3PNEFADM-FS}
\end{eqnarray}
We can see the final expression for the energy flux [Eq.~\eqref{3PNEFADM-FS}] is independent of the arbitrary length scale $z_0$. Similarly by using Eqs.~\eqref{Fhered-FS} and \eqref{EFSH-FS} [\eqref{EFMH-FS}], one can find the complete 3PN expression for energy flux in SH [MH] coordinates.
Given the total energy flux as a function of the separation between the two objects at any instant the total energy radiated during the infall can be computed as,
\begin{equation}
\Delta{E}=-\int_{z_f}^{z_i} \left({d{\cal E} \over dt}\right)\,{dz\over {\dot z}}\,,
\label{eq:deltaEFS}
\end{equation} 
where $z_i$ and $z_f$ are the initial and final separation between the two objects under head-on infall. Inserting $s=z/z_i$ and $\gamma=G\,m/c^2\,z$ back in Eq.~(\ref{3PNEFADM-FS}) and then using it along with $\dot{z}$ in ADM coordinates in Eq.~(\ref{eq:deltaEFS}) one can compute the total energy radiated during the radial infall of the two objects from a initial separation $z_i$ to a final separation $z_f$. Since Eq.~(\ref{3PNEFADM-FS}) involves some integrals which can only be evaluated numerically, we use the NIntegrate option  inbuilt in {\it Mathematica} to compute the total radiated energy during the process of infall. On the other hand for the case of infall from infinity, since we have computed the energy flux as a function of the separation between the two objects in closed form we shall provide 3PN expression for the total energy radiated during the radial infall from $z_i=\infty$ to the final separation $z_f$ in Sec.~\ref{completefluxIS}, however we wish to plot the curves corresponding to the limit $z_i=\infty$ with those corresponding to the case of infall from a finite distance for comparing the results.
\subsection{Case II: Infall from infinity}
\label{completefluxIS}
For this case the complete 3PN expression for energy flux in ADM coordinates  can be obtained by adding hereditary part (Eq.~\eqref{Fhered-IS}) and instantaneous part (Eq.~\eqref{EFADM-IS}) of energy flux and we have,
\begin{eqnarray}
\left( {d{\cal E} \over dt}\right)_{\rm ADM} 
&=& { 16 \over 15}\frac{c^5}{G}\nu^2 \gamma^5
\left \{ 1  +\left [- {43 \over 7} +{111 \over 14}\,\nu
\right]{\gamma}+\sqrt{2}\left[-
\frac{71}{6}-\frac{5}{\sqrt{3}}\,\pi
-5\,\log\left({2 \over 3}\,\gamma \right)
\right]{\gamma}^{3/2}
+ \left [-\frac{4643}{108}-\frac{713}{36}\,\nu+{112 \over 3}\, \nu ^2\right ]
{\gamma^2}\right.\nonumber \\ & & \left.
+\sqrt {2}\left[-\frac{6935}{84}+{2539\over 14\,\sqrt {3}}\,\pi+\left(
-\frac{83953}{840}-\frac{801\, \sqrt{3}}{28}\,\pi\right)\nu
+\left({691 \over 14}
-\frac{1507}{28}\,\nu\right)\,\log \left({2 \over 3}\,\gamma \right)
\right]{\gamma^{5/2}}\right.\nonumber \\ & & \left. 
+\left[\frac{363674263}{582120}
+\frac{9433}{42 \sqrt{3}}\,\pi+{145 \over 6}\,\pi^2
+\left(-\frac{5467459}{5544}-\frac{1125}{32}\,\pi^2\right)\nu+\frac{1865377}{5544}\,\nu^2\right.\right.\nonumber\\ & &\left.\left.+\frac{231520}{2079}\,\nu^3+\left(\frac{9433}{42}
+\frac{65\, \pi}{\sqrt{3}}\right)\,\log \left({2 \over 3}\,\gamma \right)
+\frac{65}{2} \log ^2\left({2 \over 3}\,\gamma\right)
+80 \psi^{(1)}\left(\frac{11}{3}\right)
\right]
{\gamma^3} \right \}\,.
\label{3PNEFADM-IS}
\end{eqnarray}
We can now see the final expression for the energy flux [Eq.~\eqref{3PNEFADM-IS}] is independent of the arbitrary length scale $z_0$. Similarly by employing Eq.~\eqref{Fhered-IS} with Eq.~\eqref{EFSH-IS} and Eq.~\eqref{EFMH-IS} one finds the complete 3PN expression for energy flux in SH and MH coordinates, respectively.

Given total energy flux as a function of the separation between the two objects at any instant the total energy radiated during the infall can be computed as
\begin{equation}
\Delta{E}=-\int_{z_f}^{+\infty} \left({d{\cal E} \over dt}\right)\,{dz\over {\dot z}}\,.
\label{eq:deltaEIS}
\end{equation}   
Using the expression for energy flux in ADM coordinates given by Eq.~\eqref{3PNEFADM-IS} and $\dot {z}$ in ADM coordinates given by Eq.~\eqref{zdot-square-ADM-IS} in the above we get the 3PN expression for total energy radiated due to head-on infall of two compact objects from infinity to a final separation of $z_f$ as 
\begin{eqnarray}
\Delta{E}_{\rm ADM}&=&{16\,\sqrt {2}\over 105}{\nu^2\, m\,c^2\,{\gamma_f}}^{7/2}\left\{1+\left[-{17 \over 6}+{187\over 36}\,\nu\right]\,{\gamma_f}
+{1\over \sqrt {2}}\left[-{91\over 6}-{7\over \sqrt {3}}\,\pi-7\,\log \left({2\over 3}\,\gamma_f\right)\right]\,{\gamma_f}^{3/2}\right.\nonumber \\ & & \left.
+\left[-\frac{10508}{297}+\frac{2191}{396}\,\nu+\frac{18323}{1056}\,\nu^2\right]{\gamma_f}^2+{1\over \sqrt {2}}\left[-138+\frac{197}{\sqrt{3}}\,\pi+\left(-\frac{5443}{60}-\frac{557}{6\sqrt{3}}\,\pi\right){\nu}\right.\right.\nonumber\\ & &\left.\left.+\left(43-{111\over 2}\,\nu\right)\,\log \left({2\over3}\,{\gamma_f}\right)\right]{\gamma_f}^{5/2}+\left[\frac{1183646333}{4684680}+\frac{9013}{78\sqrt{3}}\,\pi+\frac{1015}{78}\,\pi^2\right.\right.\nonumber\\ & &\left.\left.+\left(-\frac{32327629}{61776}-\frac{15883}{832}\,\pi^2\right){\nu}
+\frac{20519431}{82368}\,\nu ^2+\frac{17017307}{494208}\,\nu ^3+\left({9013\over 78}+{35\over \sqrt {3}}\,\pi\right)\,\log \left({2\over 3}\,\gamma_f\right)\right.\right.\nonumber\\ & &\left.\left.+{35\over 2}\,\log^2\left({2\over 3}\,\gamma_f\right)+{560\over 13}\,\psi ^{(1)}\left(\frac{11}{3}\right)\right]\,{\gamma_f}^3\right\}\,,
\label{eq:deltaEADMIS}
\end{eqnarray}
where $\gamma_f=G\,m/c^2 z_f$. 
\section{Discussions and Conclusion}
\label{discussion}
Having listed the complete 3PN expressions for the GW energy flux [Eq.~\eqref{3PNEFADM-FS} and \eqref{3PNEFADM-IS}] in ADM coordinates, in this final section
we examine its general behavior as a function of the separation between the two objects under the radial infall.
\begin{figure}[t]
\includegraphics[width=0.32\textwidth,angle=0]{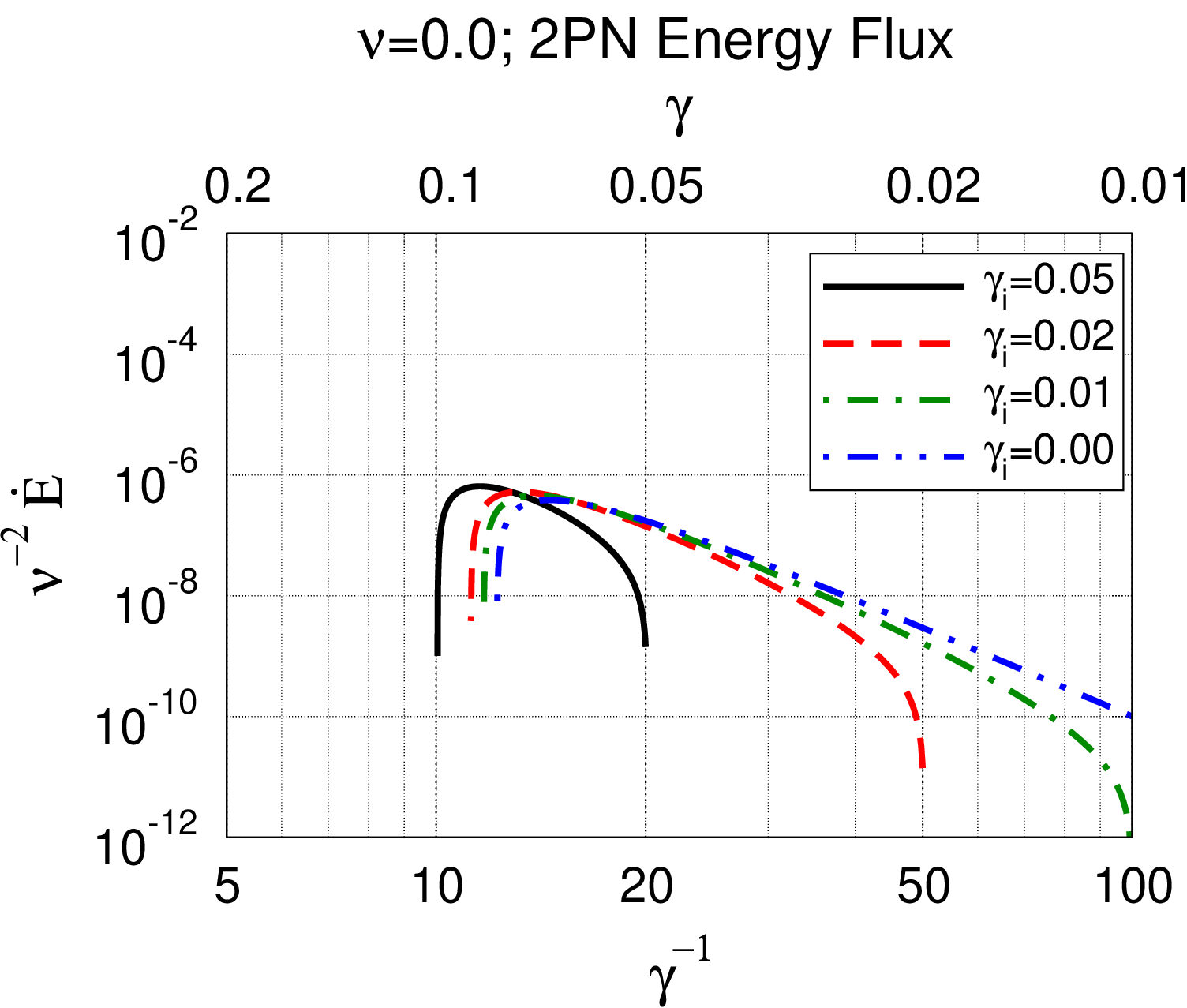}
\hskip 0.2cm
\includegraphics[width=0.32\textwidth,angle=0]{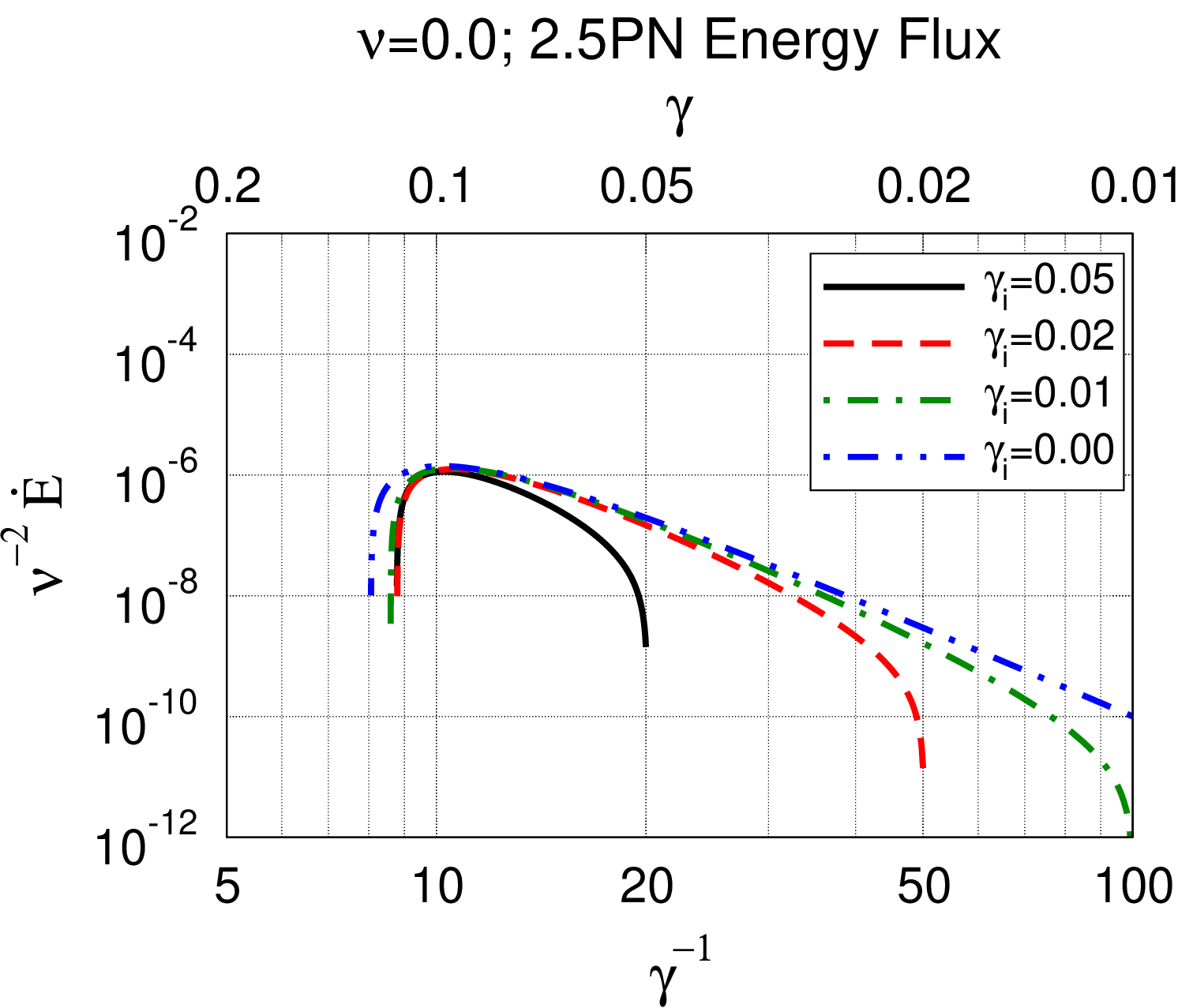}
\hskip 0.2cm
\includegraphics[width=0.32\textwidth,angle=0]{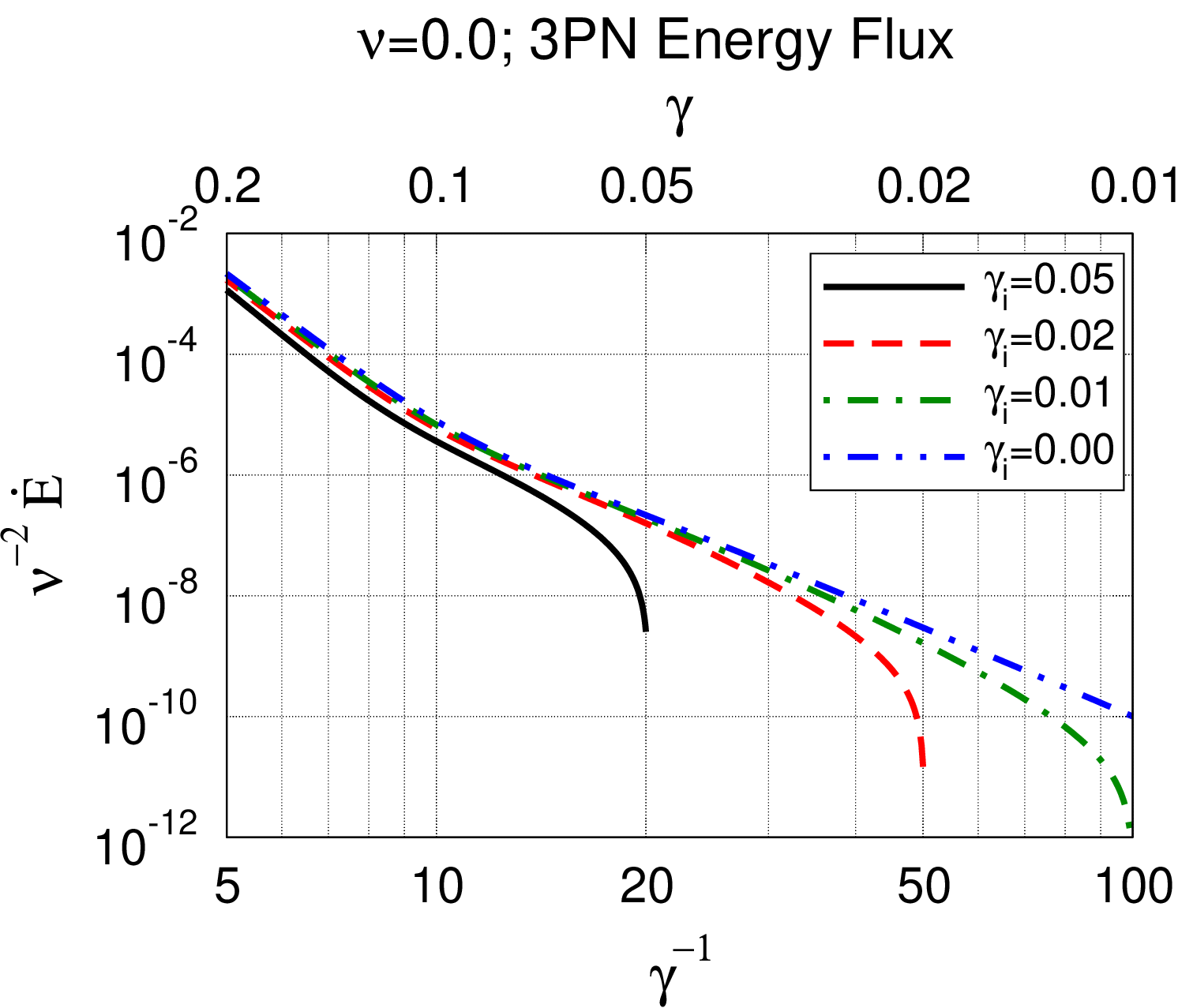}
\vskip 0.2cm
\includegraphics[width=0.32\textwidth,angle=0]{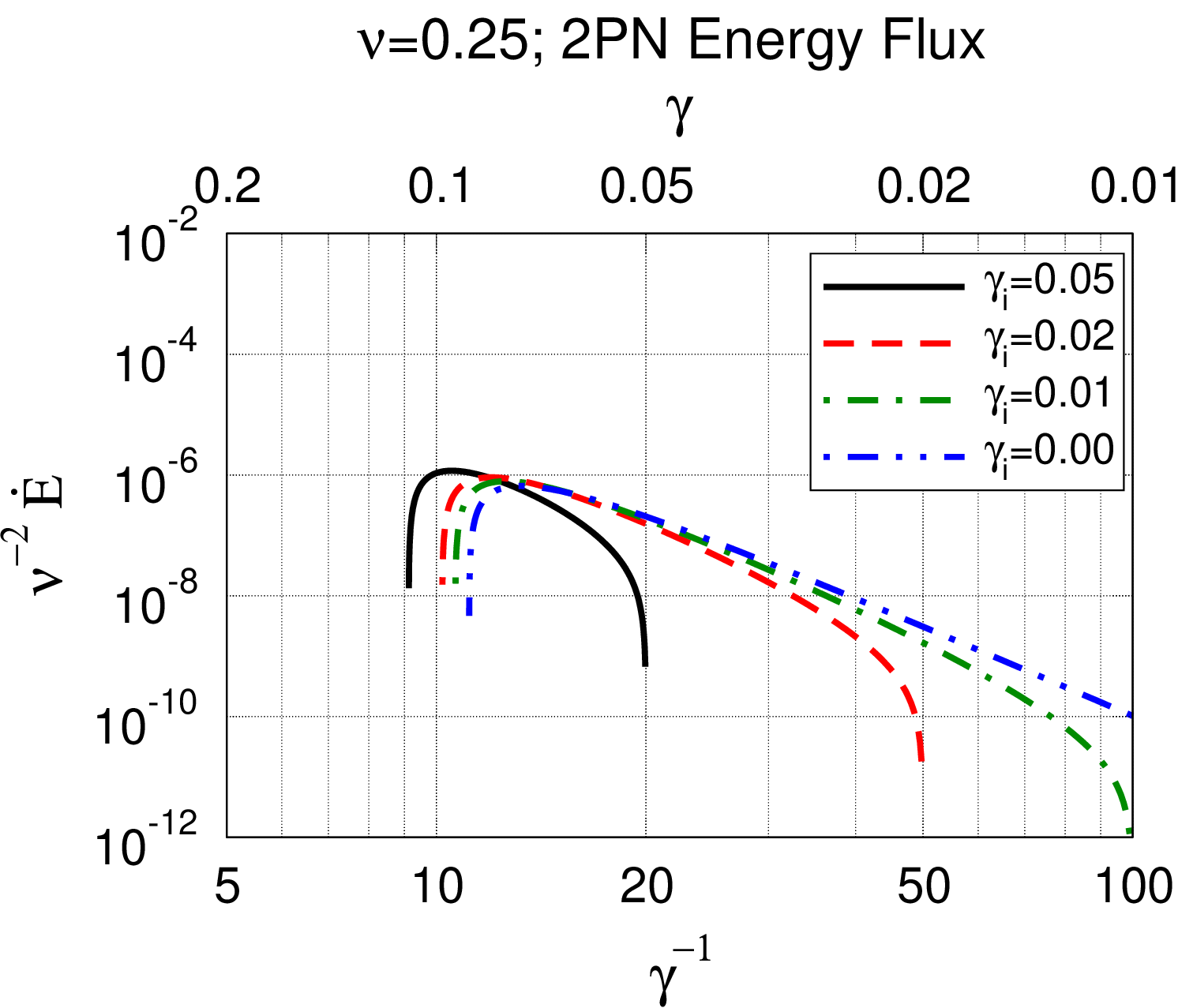}
\hskip 0.2cm
\includegraphics[width=0.32\textwidth,angle=0]{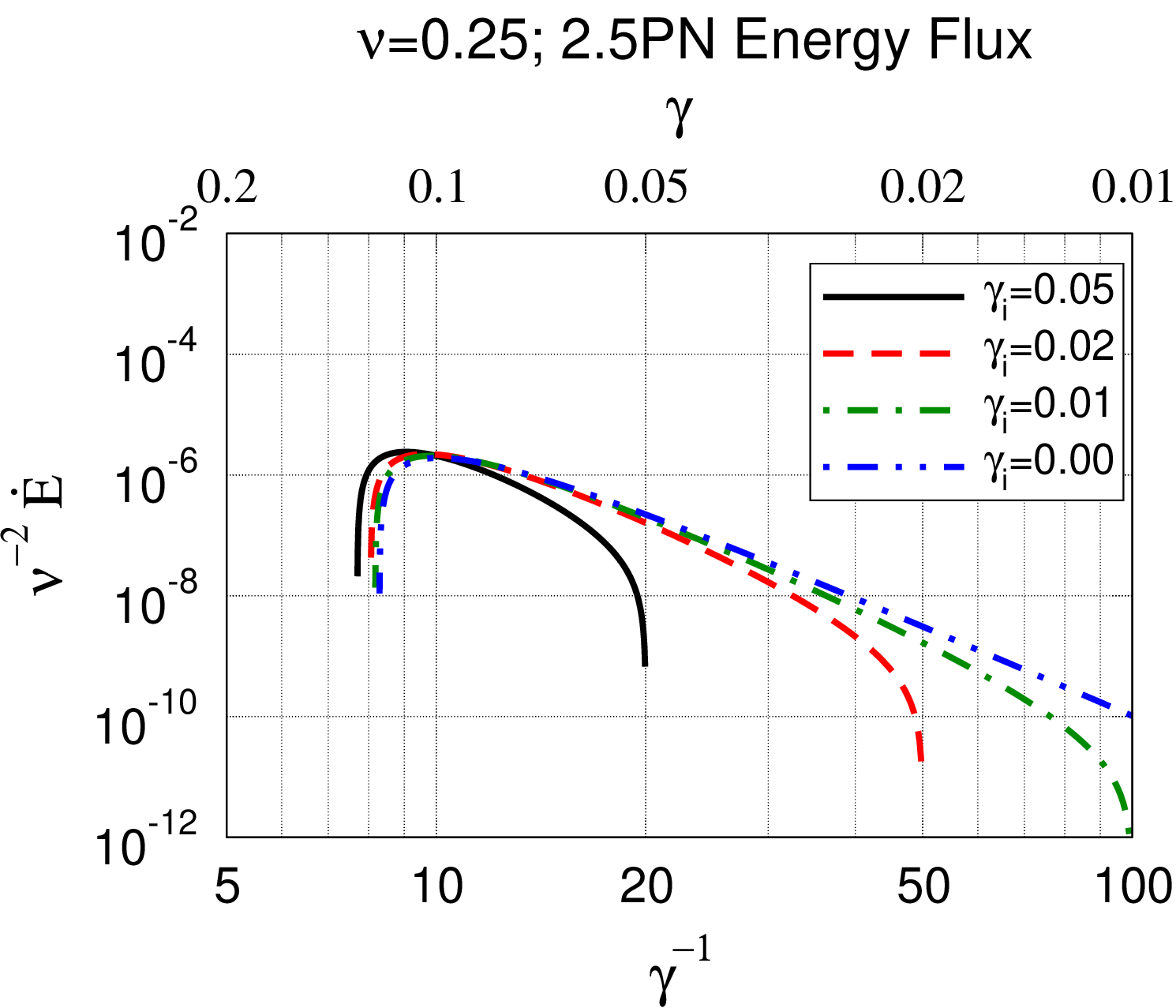}
\hskip 0.2cm
\includegraphics[width=0.32\textwidth,angle=0]{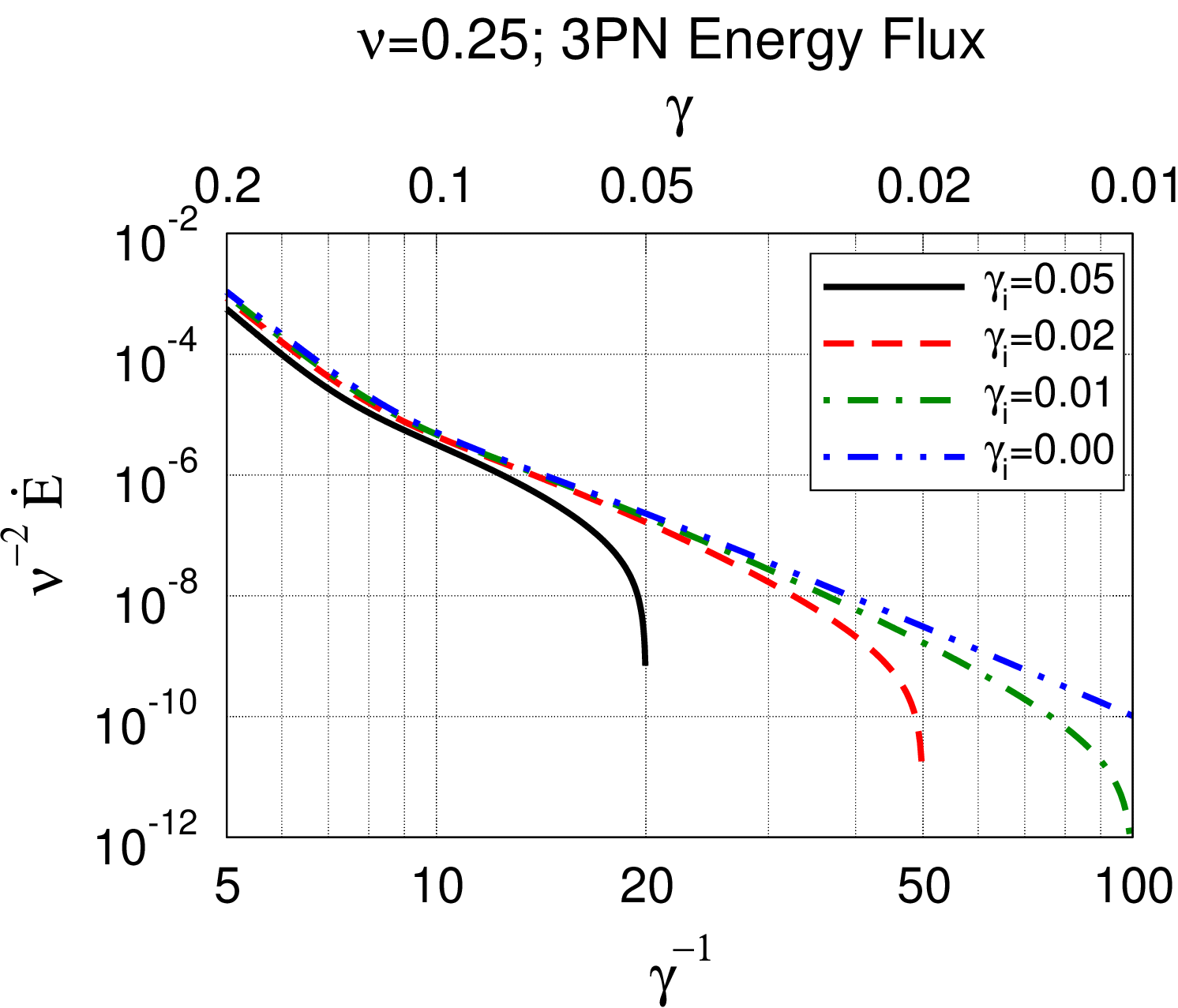}
\caption{Energy flux in ADM coordinates, in units of $\nu^2$ (where $\nu$ is the symmetric mass ratio of the binary), as a function of the parameter $\gamma=G\,m/c^2\,z$ for the head-on situation, for four different initial separations characterized by the parameter $\gamma_i=G\,m/c^2\,z_i$: $\gamma_i=0.05$, $\gamma_i=0.02$, $\gamma_i=0.01$ and $\gamma_i=0.0$ (infinite initial separation limit) which correspond to the situations when the initial separation $z_i$ between the two objects is 20\,$G\,m/c^2$, 50\,$G\,m/c^2$, 100\,$G\,m/c^2$ and $\infty$ respectively. Curves in the top panels correspond to the value of $\nu=0$ (test-body limit) while those in the bottom panels correspond to the value of $\nu=0.25$ (equal-mass case). Left, middle and right panels in both top and bottom panels correspond to the 2PN, 2.5PN, and 3PN accurate expressions for energy flux. The values given on the y-axis have been scaled by the factor $c^5/G=3.63\times10^{52}$\,$\rm{joules}$-$\rm{sec^{-1}}$. The labels on the x-axis and alternative x-axis corresponds to inverse of the parameter $\gamma$ (which is the separation between the objects under radial infall at any instant in units of $G\,m/c^2$) and the values of the PN parameter $\gamma$, respectively.}
\label{GWEFFS}
\end{figure}
Figure~\ref{GWEFFS} shows the variation of the energy flux, in units of $\nu^2$ scaled by a factor $c^5/G=3.63\times10^{52}$\,$\rm{joules}$-$\rm{sec^{-1}}$,
as a function of the parameter $\gamma$  in ADM coordinates 
(recall $\gamma=G\,m/c^2\,z$
where $z$ is the instantaneous separation between the two compact objects falling radially towards each other). 
Each panel in Fig.~\ref{GWEFFS} shows a comparison between the energy flux emitted as a function of the parameter $\gamma$ for different initial separations including the limiting case
of infinite initial separation as well. 
In each panel curves corresponding to different initial separations (characterized by the parameter $\gamma_i=G\,m/c^2z_i$) have been plotted for $\gamma_i= 0.05, 0.02, 0.01,$ and $ 0.0 $
and correspond to the situation when the initial separation $z_i$ between the two objects  is 20\,$G\,m/c^2$, 50\,$G\,m/c^2$, 100\,$G\,m/c^2$ and $\infty$ respectively. 
Curves in the top panels correspond to $\nu=0$ (test-body limit) while those in the bottom panels correspond to $\nu=0.25$ (equal-mass case). It is obvious from the figure that the curves in each panel approach each other with increasing $\gamma$ i.e.
when the separation between the two objects decreases. 
This feature can be understood by recalling that since $s=z/z_i=\gamma_i/\gamma$, for a fixed $z_i$,  the finite-separation corrections in powers of  $s$ become progressively
less important as the bodies approach each other (small $z$). The finite-separation effects, important when the objects
are far apart, are less significant at closer separation and the curves for the  energy flux approach each other. 
 
Figure~\ref{GWEFFS} also compares the results that would be obtained using the 2PN, 2.5PN and 3PN accurate expression for the energy flux
and thus illustrates the improvements  arising from a more accurate expression for the energy flux. It is clear from  Fig.~\ref{GWEFFS} that the energy flux emitted at any instant monotonically increases as the separation between the objects under the infall decreases (with increasing $\gamma$) as generally expected. However from  Fig.~\ref{GWEFFS} we see that after a certain maximum value of the parameter $\gamma$  in the 2PN and 2.5PN cases the curves show a turnover and start to decrease. This is an indication of the fact that the PN approximation is no longer valid beyond this value of $\gamma$. 
It should be noted  that the value of $\gamma$ where this  happens depends upon the choice of the initial separation between the two objects, the PN accuracy of the expression for the energy flux 
and the symmetric mass ratio of the binary.

\begin{figure}[t]
\includegraphics[width=0.32\textwidth,angle=0]{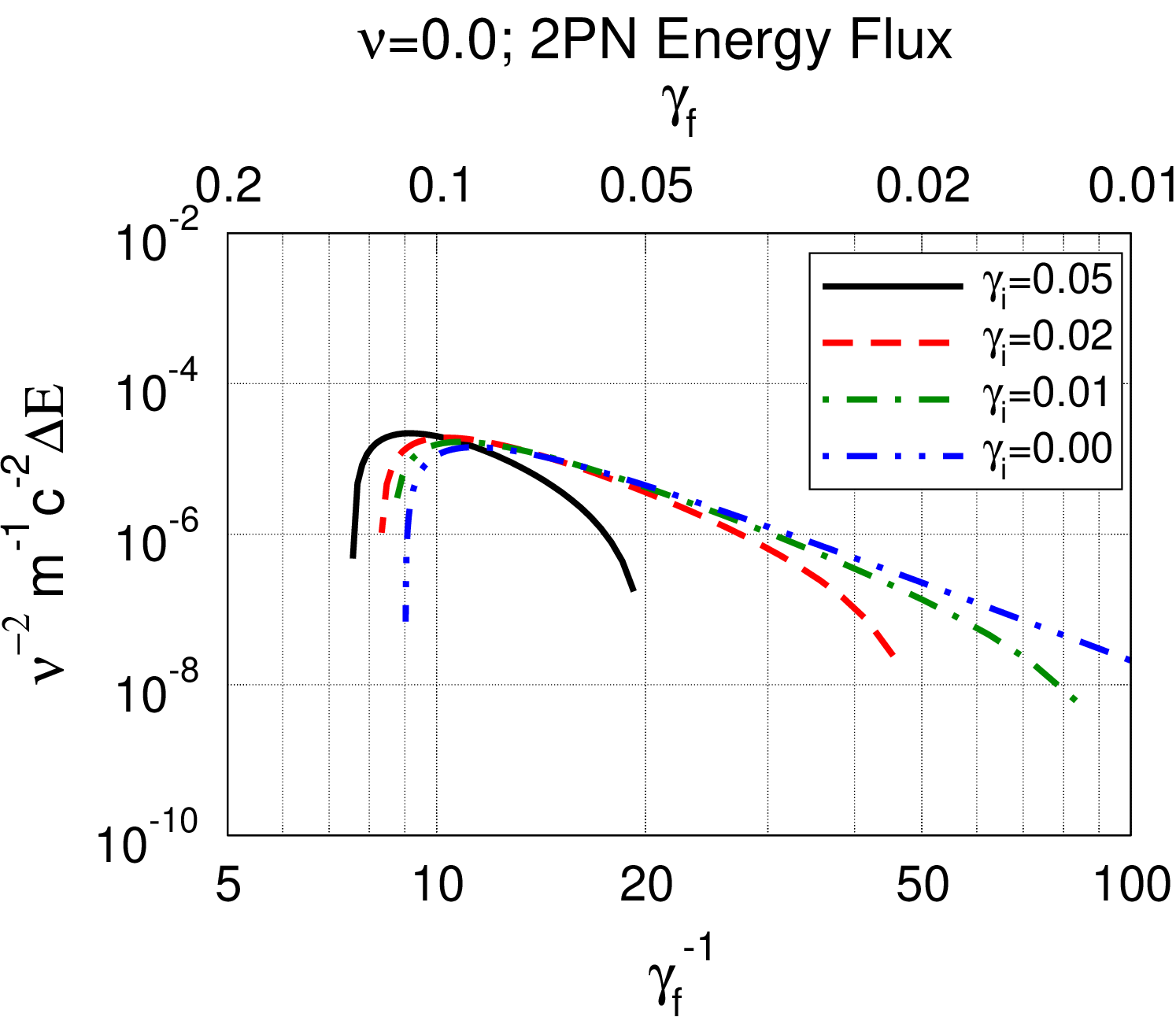}
\hskip 0.2cm
\includegraphics[width=0.32\textwidth,angle=0]{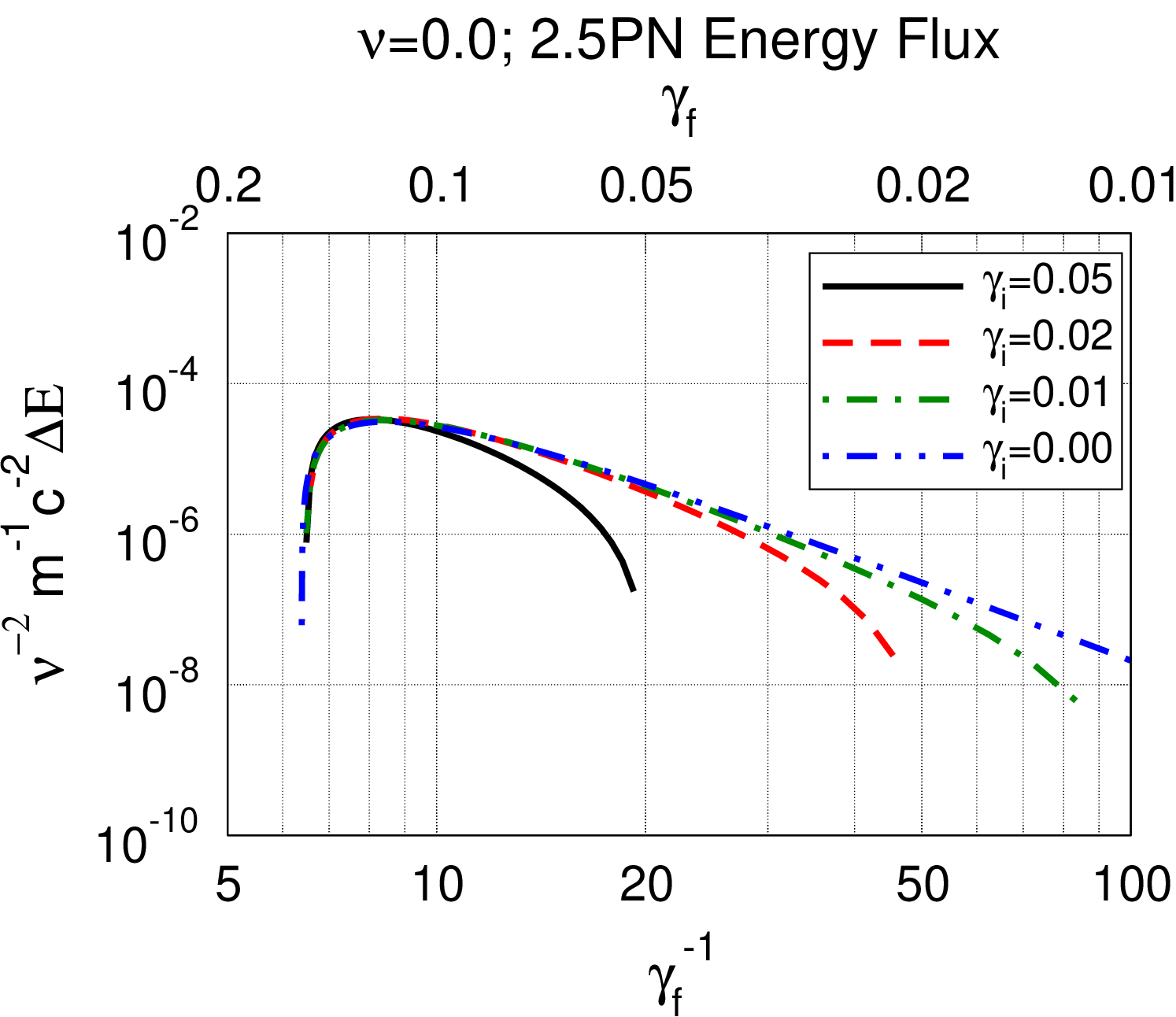}
\hskip 0.2cm
\includegraphics[width=0.32\textwidth,angle=0]{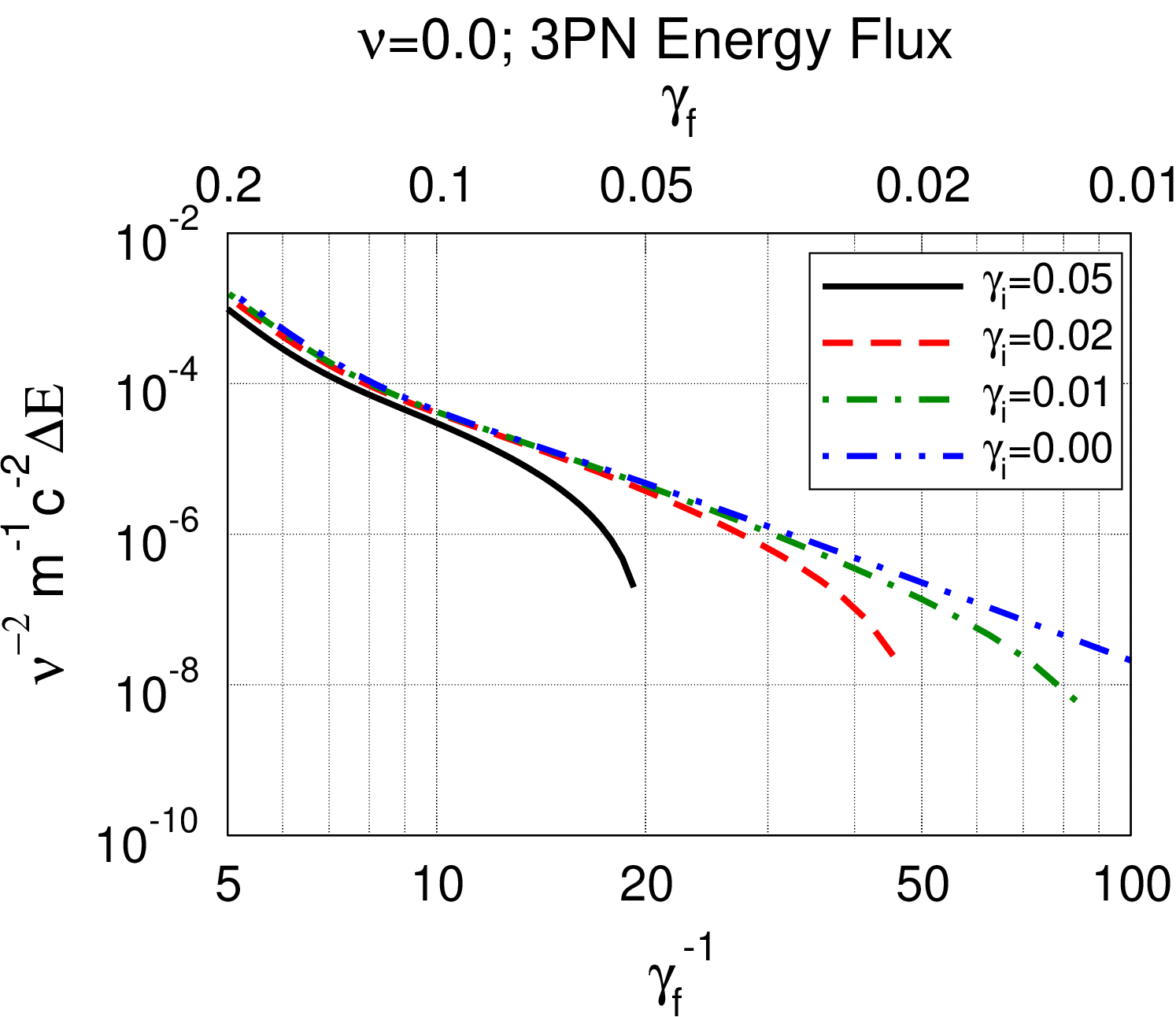}
\vskip 0.2cm
\includegraphics[width=0.32\textwidth,angle=0]{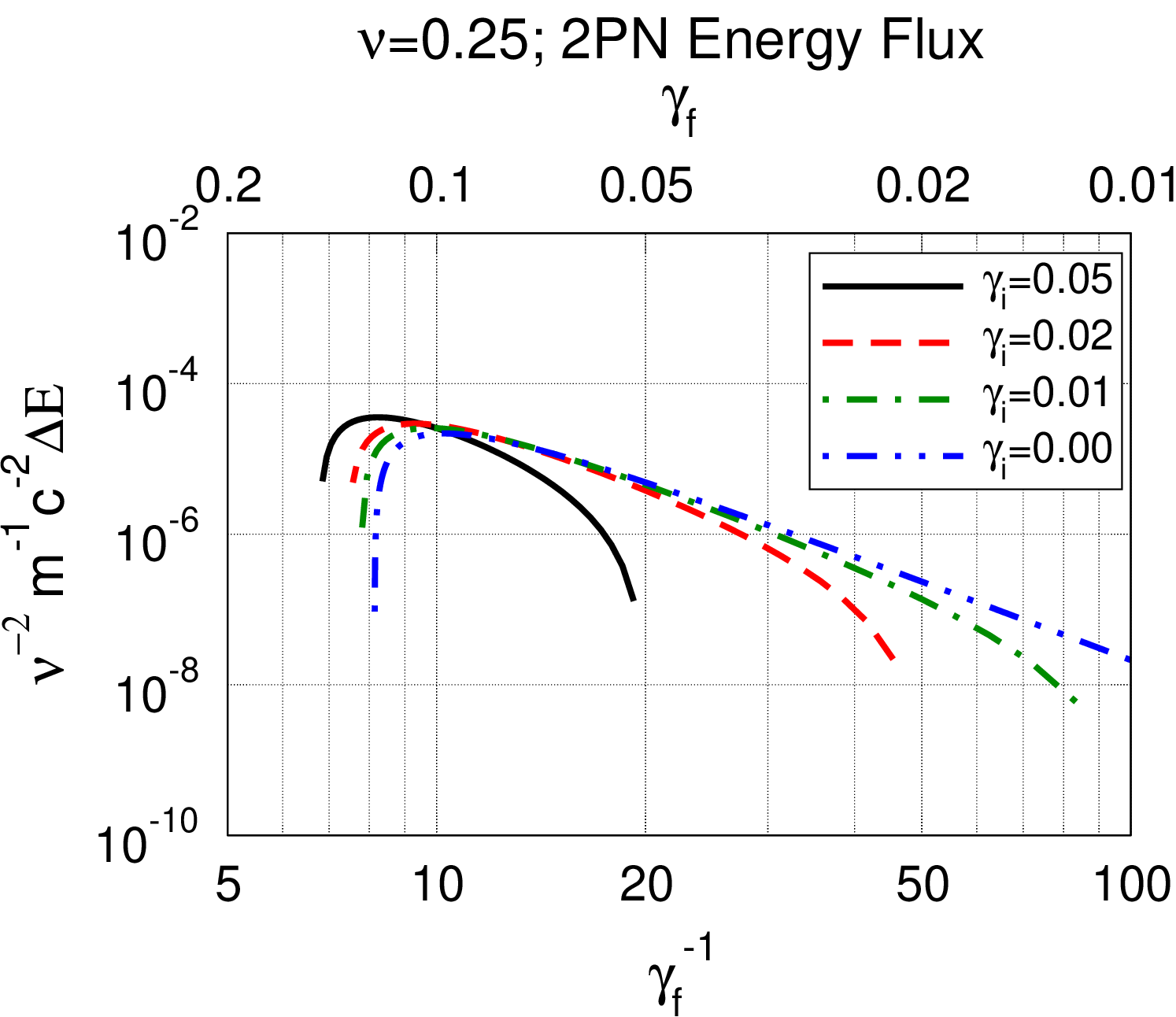}
\hskip 0.2cm
\includegraphics[width=0.32\textwidth,angle=0]{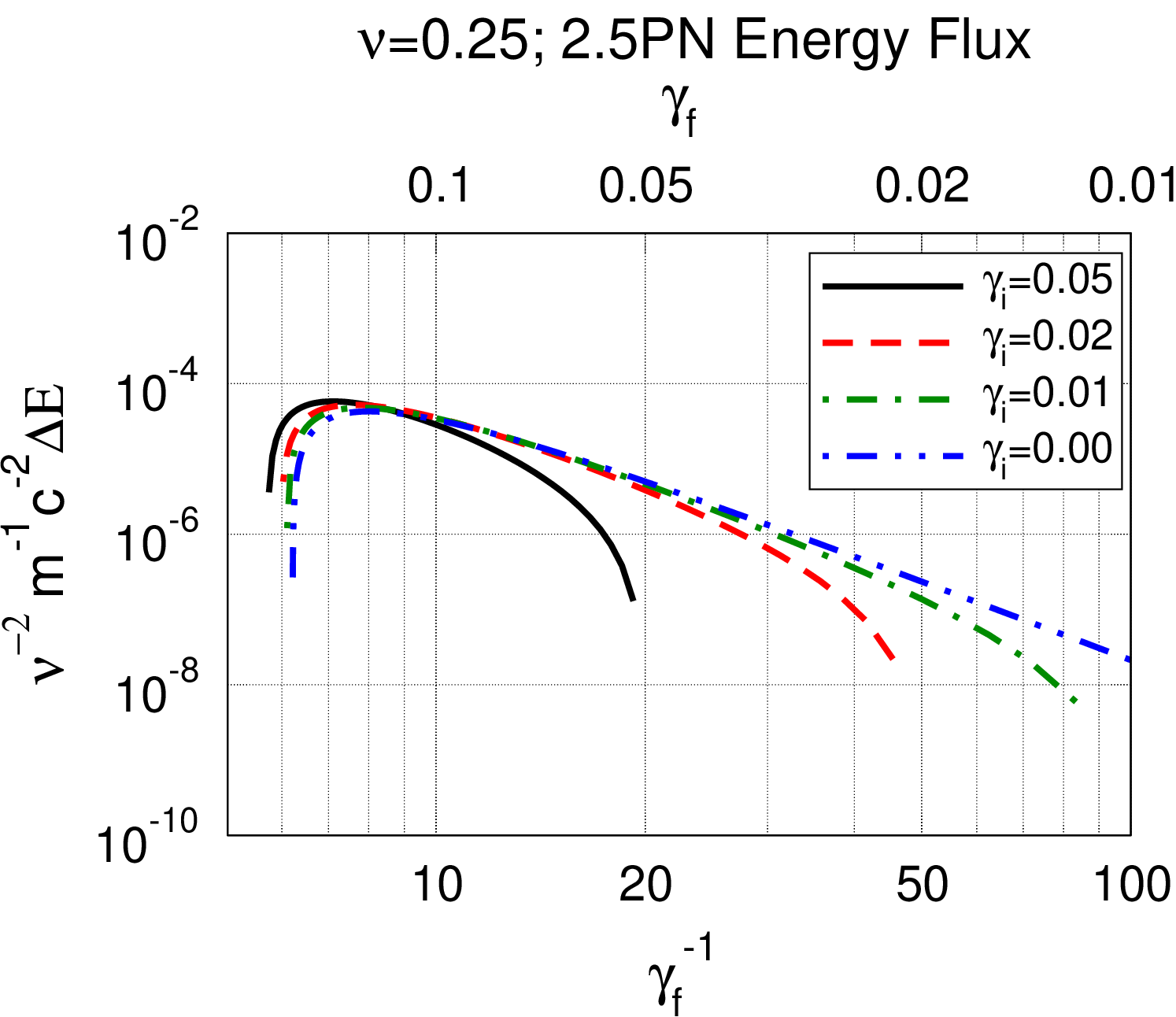}
\hskip 0.2cm
\includegraphics[width=0.32\textwidth,angle=0]{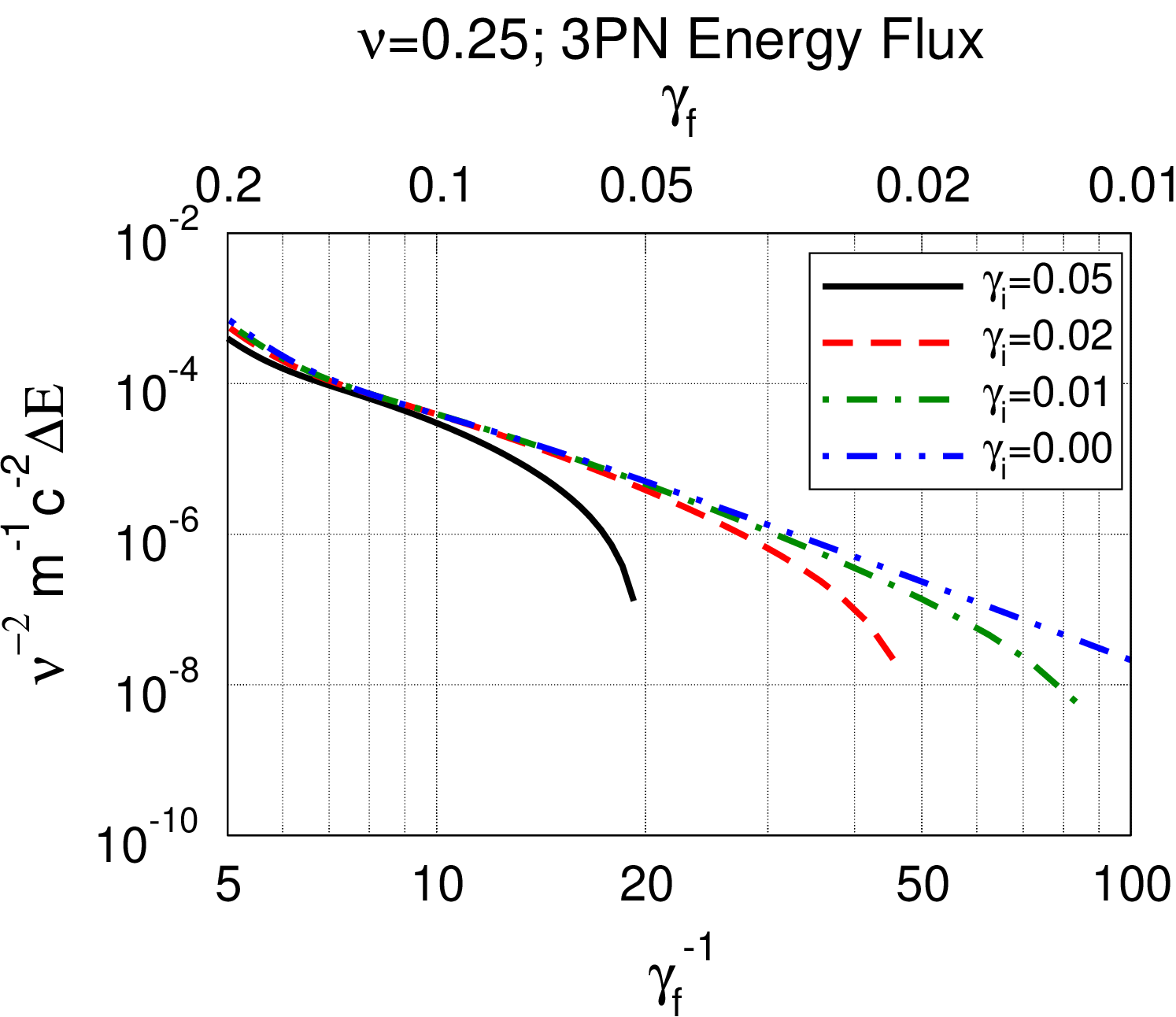}
\caption{Similar to Fig.~\ref{GWEFFS} but  for total energy radiated
 in units of $\nu^2\,c^2\,m$,  during the head-on infall of two compact objects from a initial separation $z_i$ in ADM coordinates 
(related to the parameter $\gamma_i$) to a final separation of $z_f$ (corresponding parameter $\gamma_f=G\,m/c^2\,z_f$).}
\label{Fig:deltaEADMFS}
\end{figure}

Finally, Fig.~\ref{Fig:deltaEADMFS} shows the total energy radiated [as discussed in the previous section for the finite initial separation case it has to be computed numerically 
using Eq.~\eqref{eq:deltaEFS} but for the infinite initial separation case it is given by Eq.~\eqref{eq:deltaEIS}] during a radial infall from initial separation $z_i$ 
(characterized by the parameter $\gamma_i=G\,m/c^2\,z_i$) to a final separation $z_f$ (characterized by the parameter $\gamma_f=G\,m/c^2\,z$). 
Similar to Fig.~\ref{GWEFFS} in Fig.~\ref{Fig:deltaEADMFS} we study the effect of using different PN-accurate expressions for energy flux and also the effect of assuming different initial separations 
in the problem. It is evident from each panel of the Fig.~\ref{Fig:deltaEADMFS} that as $\gamma_f$ ($z_f$) increases (decreases) all curves approach each other which 
implies  that most of the contribution comes from the late stages of the infall. 
It is evident from the plots in Fig.~\ref{Fig:deltaEADMFS} that only beyond a certain minimum separation between the two objects (under the infall) the estimates of energy radiated
using PN expressions are valid.  The 2PN, 2.5PN and 3PN estimates of the total energy radiated during the radial infall (from infinity) of two  equal mass compact objects  is of the order of 
$2.2\times10^{-5}$, $4.3\times10^{-5} {\rm and}\, 7.4\times 10^{-5}$ respectively. In the test particle limit
The corresponding 2PN, 2.5PN and 3PN accurate results for total energy radiated in the test particle limit 
are of the order of $1.4\times 10^{-5}, 3.1\times 10^{-5} {\rm and}\, 8.5\times10^{-5}$ respectively. 
 Unlike the 2PN and 2.5PN cases where the breakdown of the PN approximation is explicit in the turnover, the 3PN approximation does not show any sharp turnover.
As a consequence the value quoted for the maximum energy radiated in the 3PN case is a bit arbitrary  and corresponds to the value at the point where
the 2.5PN approximation breaks down.    
From the Fig.~\ref{Fig:deltaEADMFS} one can infer that the energy radiated in the process of head-on infall  for the finite separation cases ($\gamma_i=0.05, 0.02, 0.01$) 
is of the same order as in infinite initial separation case ($\gamma_i=0$).
It is evident from the above discussion that the 3PN estimates of the peak luminosities and the energy loss in form gravitational radiation during the infall between the initial ($z_i$) and a final point ($z_f$) will not only be more than the estimates of the same using a less accurate expressions (2PN and 2.5PN accurate) but also they are valid till later stages of the infall and thus allows one to compare the results obtained using numerical relativity within the range in which PN approximations are valid.     
\appendix
\section{Calculation of $\delta_{\rm (SH\rightarrow ADM)}E$}
\label{deltaESHtoADM}
General expression for energy $E$ in CM frame associated with SH coordinate system is given in terms of the natural variables; $r$, $v$ and $\dot {r}$~\cite{Blanchet:2002mb}. Noticing this functional dependence and the fact that it is a scalar quantity we expect that under a transformation ($r' \rightarrow r+\delta {r}$, $v' \rightarrow v+\delta {v}$, $\dot {r}' \rightarrow \dot {r}+\delta {\dot {r}}$ ) this would transform in CM frame as
\begin{equation}
E'=E+\delta {E}
\end{equation}
Or equivalently for transformations between SH and ADM coordinate systems,
\begin{equation}
E_{\rm ADM}=E_{\rm SH}+\delta_{\rm (SH\rightarrow ADM)}{E}\,,
\end{equation} 
where,
\begin{equation} 
\delta_{\rm (SH\rightarrow ADM)} {E}=\delta {r}{\partial E \over \partial r}+\delta {v}{\partial E \over \partial v}+\delta {\dot r}{\partial E \over \partial \dot r}\,.  
\end{equation}
The shifts in the variables $r$, $v$ and $\dot {r}$ connecting ADM and SH coordinates are given by Eq.~(6.10) of~\cite{ABIQ07} and the expression for CM energy $E_{\rm SH}$ for general orbits is given by Eq.~(4.8) of~\cite{Blanchet:2002mb}. Having all inputs we now can write the shift $\delta_{\rm (ADM \rightarrow SH)} {E}$ for general orbits which reads as
\begin{eqnarray}
\delta_{\rm (SH\rightarrow ADM)} {E}&=&{G\,m^2\,\nu \over c^4\,r}\biggl[-{13\over 8}\,\nu\,v^4+{5 \over 4}\,\nu\,v^2\,\dot {r}^2+{G\,m \over r}\,v^2\left({1 \over 4}+{47\over 8}\,\nu\right)\nonumber \\
&+&{3 \over 8}\,\nu\,\dot {r}^4
+{G\, m \over r}\,\dot {r}^2\left(-{1\over 2}-{57\over 8}\,\nu\right)+{G^2\,m^2 \over r^2}\left({1\over 4}+3\,\nu\right)
\biggr]\nonumber\\
&+&{G\,m^2\,\nu \over c^6\,r}\biggl[v^6\left(-{65\over 16}\,\nu+{179 \over 16}\,\nu^2\right)+\dot {r}^2\,v^4\left({61\over 16}\,\nu 
-{165 \over 16}\,\nu^2\right)\nonumber \\&+&{G\,m \over r}\,v^4\left({3 \over 8}+{7 \over 8}\,\nu-{481 \over 16}\,\nu^2\right)+\dot {r}^4\,v^2\left({9\over 16}\,\nu -{39\over 16}\,\nu^2\right)+{G\,m \over r}\,\dot {r}^2\,v^2\left(-{3 \over 4}-{131\over 16}\,\nu +{641 \over 16}\,\nu^2\right)\nonumber \\
&+&{G^2\,m^2\over r^2}\,v^2\left({3\over 8}+\left[{26167\over 1680}-{21\,\pi^2\over 32}+{22\over 3}\,\log \left({r\over r'_0}\right)\right]\nu+{37\over 8}\,\nu^2\right)+\dot {r}^6\left(-{5\over 16}\,\nu+{25\over 16}\,\nu^2\right)\nonumber\\
&+&{G\,m\over r}\,\dot {r}^4\left({671\over 48}\,\nu-{6\,\nu^2}\right)+{G^2\,m^2\over r^2}\,\dot {r}^2\left(-{3\over 2}+\left[{6479\over 1680}+{63\,\pi^2\over 32}-22\,\log \left({r \over r'_0}\right)\right]\nu-{23\over 2}\,\nu^2\right)\nonumber \\
&+&{G^3\,m^3\over r^3}\left(-{1\over 4}+\left[-{3613\over 280}-{21\,\pi^2\over 32}+{22\over 3}\log \left({r\over r'_0}\right)\right]\nu\right)\biggr]
\label{deltaSHtoADMGO}
\end{eqnarray}
It is easy to see that when restrictions given by Eq.~\eqref{GOtoHOC} are imposed, the above expression reduces to the form given by Eq.~\eqref{EnSHtoADM}.  
\begin{acknowledgments}
We thank L. Blanchet for useful suggestions on the manuscript.
\end{acknowledgments}
\bibliography{ref-list}
\end{document}